\documentclass[11pt]{article}

\usepackage{graphicx}
\usepackage{amsfonts}
\usepackage{amssymb}
\usepackage{amsthm}
\usepackage{amsmath}
\usepackage{multirow}

\newcommand{\newc}{\newcommand}
\newc{\ra}{\rightarrow}
\newc{\lra}{\leftrightarrow}
\newc{\be}{\begin{equation}}
\newc{\ee}{\end{equation}}
\newc{\bs}{\begin{split}}
\newc{\es}{\end{split}}
\newc{\bg}{\begin{gathered}}
\newc{\eg}{\end{gathered}}
\newc{\ba}{\begin{eqnarray}}
\newc{\ea}{\end{eqnarray}}
\newc{\ov}{\overline}
\newc{\pa}{\partial}
\newc{\D}{\Delta}

\newc{\nn}{\nonumber}

\begin{document}

\begin{titlepage}

\vspace*{0.7cm}

\begin{center}
{
\bf\LARGE
Discrete Family Symmetry from F-Theory GUTs}
\\[12mm]
Athanasios~Karozas$^{\dagger}$
\footnote{E-mail:\texttt{akarozas@cc.uoi.gr}},
Stephen~F.~King$^{\star}$
\footnote{E-mail: \texttt{king@soton.ac.uk}},
George~K.~Leontaris$^{\dagger}$
\footnote{E-mail: \texttt{leonta@uoi.gr}},
Andrew~K.~Meadowcroft$^{\star}$
\footnote{E-mail: \texttt{am17g08@soton.ac.uk}}
\\[-2mm]

\end{center}
\vspace*{0.50cm}
\centerline{$^{\star}$ \it
School of Physics and Astronomy, University of Southampton,}
\centerline{\it
SO17 1BJ Southampton, United Kingdom }
\vspace*{0.2cm}
\centerline{$^{\dagger}$ \it
Physics Department, Theory Division, Ioannina University,}
\centerline{\it
GR-45110 Ioannina, Greece}
\vspace*{1.20cm}

\begin{abstract}
\noindent 
We consider realistic F-theory GUT models 
based on discrete family symmetries $A_4$ and $S_3$, combined with $SU(5)$ GUT,
comparing our results to existing field theory models based on these groups.
We provide an explicit calculation to support the emergence of the family symmetry 
from the discrete monodromies arising in F-theory. We work within
the spectral cover picture where in the present context the discrete symmetries
are associated to monodromies among  the roots of a five degree
polynomial and hence constitute a subgroup of the $S_5$ permutation symmetry. 
 We focus on the  cases of $A_4$ and $S_3$ subgroups, motivated  by 
 successful phenomenological models interpreting the fermion mass hierarchy 
 and in particular the neutrino data. More precisely, we study the implications on the effective field  theories 
by analysing  the relevant  discriminants and the topological properties of the polynomial coefficients,
   while we  propose a discrete version of the doublet-triplet splitting mechanism.
 \end{abstract}

 \end{titlepage}

\thispagestyle{empty}
\vfill
\newpage

\setcounter{page}{1}

\section{Introduction}

F-theory is defined on an elliptically fibered Calabi-Yau four-fold over a threefold base~\cite{Vafa:1996xn}.
In the elliptic fibration the singularities of the internal manifold are associated to the gauge symmetry. 
The basic objects in these constructions are the D7-branes which are located at the ``points'' where the fibre degenerates,
while matter fields appear at their intersections. The interesting fact in this  picture is that
the topological properties of  the internal space are converted 
to constraints on the effective field theory model in a direct manner. Moreover, in these constructions
it is possible to implement a flux mechanism which breaks the symmetry and generates 
chirality in the spectrum.

F-theory Grand Unified Theories
(F-GUTs)~\cite{Donagi:2008ca,Beasley:2008dc,Beasley:2008kw,Blumenhagen:2009yv,Heckman:2008qa,Heckman:2008es,Blumenhagen:2008zz}
represent a promising framework for addressing the flavour problem of quarks and leptons
(for reviews see \cite{Denef:2008wq,Weigand:2010wm,Heckman:2010bq,Grimm:2010ks,Leontaris:2012mh,Maharana:2012tu}).
F-GUTs are associated with D7-branes wrapping a complex surface $S$ in an elliptically fibered eight dimensional internal space. The precise gauge group is determined by the specific structure of the singular fibres over the compact surface $S$, which is strongly constrained by the Kodaira  conditions. 
The so-called ``semi-local'' approach imposes constraints from requiring that S is embedded into a local Calabi-Yau four-fold, which in practice leads to the presence of a local $E_8$ singularity \cite{Heckman:2009mn},
which is the highest non-Abelian symmetry allowed by the elliptic fibration.

  In the convenient Higgs bundle picture and in particular the spectral cover approach, 
  one may work locally by picking up a subgroup 
 of $E_8$ as the gauge group of the four-dimensional effective model while  the commutant of it with respect
 to $E_8$ is associated to the geometrical properities in the vicinity. 
 Monodromy actions, which are always present in F-theory constructions, may reduce the rank of the latter, 
 leaving intact only a subgroup of it. The remaining symmetries could be $U(1)$ factors in the
 Cartan subalgebra or some discrete symmetry.   Therefore, in these constructions GUTs are 
 always accompanied by additional symmetries which play important role
 in low energy pheomenology through the restrictions they impose on superpotential couplings.

In the above approach, 
all Yukawa couplings originate from this single point of $E_8$ enhancement. As such, we can learn about the matter and couplings of the semi-local theory by decomposing the adjoint of $E_8$ in terms of representations of the GUT group and the perpendicular gauge group.  In terms of the local picture considered so far, matter is localised on curves where the GUT brane intersects other 7-branes with extra $U(1)$ symmetries associated to them, with this matter transforming in bi-fundamental representations of the GUT group and the $U(1)$.  Yukawa couplings are then induced at points where three matter curves intersect, corresponding to a further enhancement of the gauge group.

Since $E_8$ is the highest symmetry of the elliptic fibration, the gauge symmetry of the effective model can in principle be any of the $E_8$ subgroups.  The gauge symmetry can be broken by turning on appropriate fluxes \cite{Donagi:2008kj} which at the same time generate chirality for matter fields.
The minimal scenario of $SU(5)$ GUT has been extensively studied
~\cite{Marsano:2009gv,Hayashi:2009bt,Dudas:2010zb}. Indeed only the simplest $SU(5)$ GUTs can in principle avoid exotic matter in the spectrum \cite{Beasley:2008kw}. 
However, by considering different fluxes, other models have been constructed with different GUT groups, such as $SO(10)$ 
and $E_6$~\cite{King:2010mq,Callaghan:2011jj,Antoniadis:2012yk,Callaghan:2012rv,Tatar:2012tm}. 
In particular, it is possible to achieve gauge coupling unification from $E_6$
in the presence of TeV scale exotics originating from both the matter curves and the 
bulk~\cite{Callaghan:2013kaa}.

All of the approaches mentioned so far
exploit the extra $U(1)$ symmetries as family symmetries, in order to address the quark and lepton mass hierarchies. While it is gratifying that such symmetries can arise from 
a string derived model, where the parameter space is subject to constraints from the first principles 
of the theory, the possibility of having only continuous Abelian family symmetry in F-theory represents a 
very restrictive choice. By contrast, other string theories have a rich group structure embodying both continuous as well as discrete symmetries at the same time~\cite{Ibanez:1991hv}-\cite{BerasaluceGonzalez:2012vb}. 
It may be regarded as something of a drawback of the F-theory approach 
that the family symmetry is constrained to be a product of $U(1)$ symmetries.
Indeed the results of the neutrino oscillation experiments are in agreement with an
almost maximal atmospheric mixing angle $\theta_{23}$, a large solar mixing 
 $\theta_{12}$, and a non-vanishing but smaller reactor angle $\theta_{13}$,
 all of which could be explained by an underlying non-Abelian discrete  
family symmetry (for recent reviews see for example~\cite{Altarelli:2010gt,King:2013eh,King:2014nza}).

Recently, discrete symmetries in F-theory have been considered~\cite{Antoniadis:2013joa}
on an elliptically fibered space with an $SU(5)$ GUT singularity,
where the effective theory is invariant under a more general non-Abelian finite group.
They considered all possible monodromies which induce an additional discrete (family-type) symmetry on the model. For the $SU(5)$ GUT minimal unification scenario in particular, 
the accompanying  discrete family group could be any subgroup of the $S_5$ 
permutation symmetry, and the spectral cover geometries with monodromies associated to the finite symmetries 
$S_4$, $A_4$ and their transitive subgroups, including the dihedral group $D_4$ and $Z_2\times Z_2$,
were discussed. However a detailed analysis was only presented for the $Z_2\times Z_2$ case,
while other cases such as $A_4$ were not fully developed into a realistic model.

In this paper we extend the analysis in~\cite{Antoniadis:2013joa} in order to 
construct realistic models   based on the cases $A_4$ and $S_3$, combined with $SU(5)$ GUT,
comparing our results to existing field theory models based on these groups.
We provide an explicit calculation to support the emergence of the family symmetry as 
from the discrete monodromies.  In  section  2 we start with a short description of the
basic ingredients of  F-theory model building and present the splitting of the spectral cover in 
the components  associated to the $S_4$ and $S_3$ discrete group factors.  
In section 3 we  discuss the conditions for the transition  of $S_4$ to $ A_4$ discrete 
family symmetry  ``escorting'' the  $SU(5)$ GUT  and propose a discrete version of the doublet-triplet
 splitting mechanism for $A_4$, before constructing a realistic model
which is analysed in detail. In section 4 we then analyse in detail an $S_3$ model which was not considered at all in
~\cite{Antoniadis:2013joa} and in section 5 we present our conclusions. Additional computational details are left for the
Appendices.

\newpage

\section{General Principles}

\begin{figure}[!bth]
  \centering
  \includegraphics[scale=0.5]{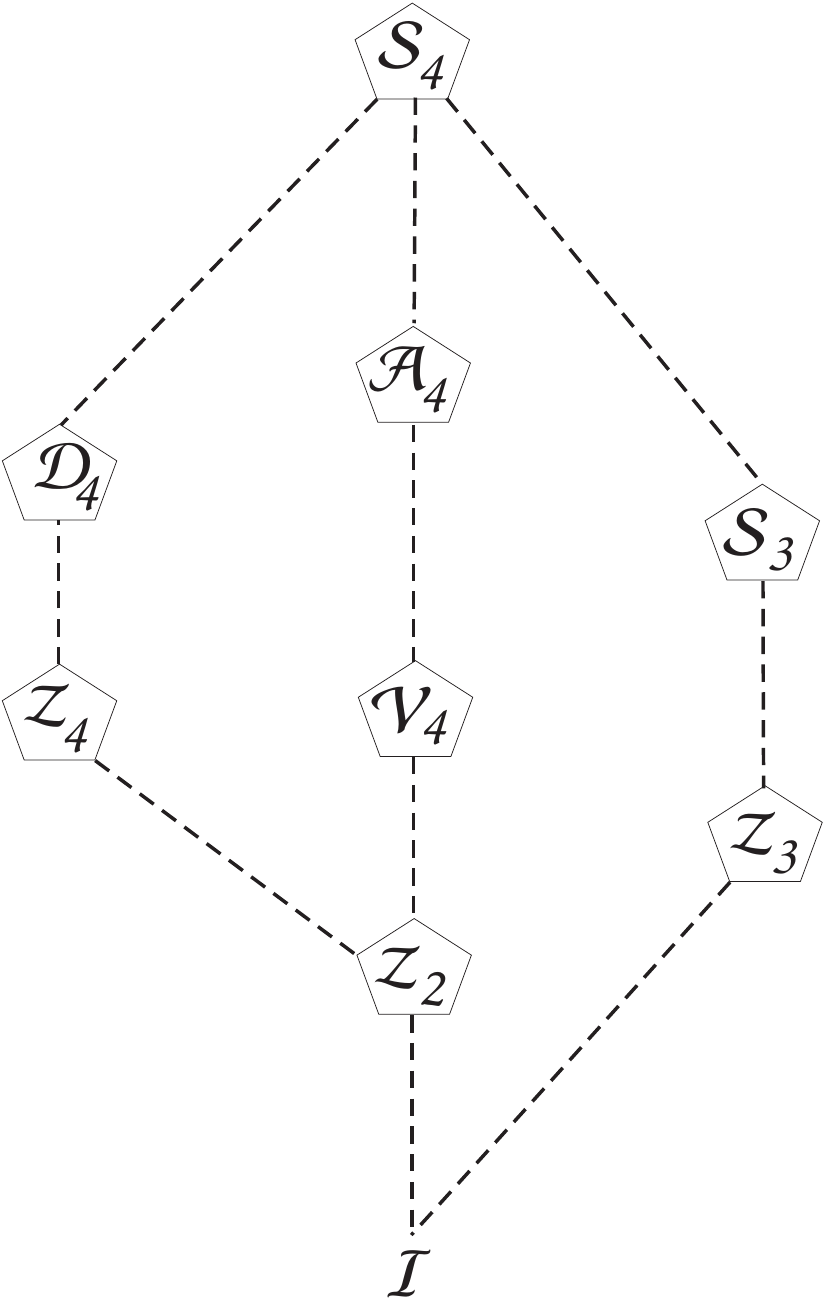}
  \caption{$S_4$ and its subgroups relevant to the present analysis.  }
  \label{S4chains}
  \end{figure}

F-theory is a non-perturbative formulation of type IIB superstring theory, emerging
 from compactifications on a Calabi-Yau fourfold which is an elliptically
fibered space over a base $B_3$ of three complex dimensions. 
Our GUT symmetry in the present work is $SU(5)$ which is associated to
a holomorphic divisor residing  inside the threefold base, $B_3$. If we designate with $z$
 the `normal' direction to this GUT surface, the divisor can be thought of
 as the zero limit of the holomorphic section $z$ in $B_3$, i.e. at $z\to 0$. The fibration 
 is described by the Weierstrass equation
 \[ y^2=x^3+f(z)x+g(z)\,,\]
where $f(z), g(z)$ are eighth and twelveth degree polynomials respectively.  The singularities
of the fiber are determined by the zeroes of the discriminant $\Delta =4 f^3+27 g^2$ 
and are associated to non-Abelian gauge groups.  For a smooth Weierstrass model 
they have been classified by  Kodaira   and in the case of F-theory these 
have been used to describe the non-Abelian gauge group.\footnote{For mathematical background see for example ref~\cite{Nakayama88}}
Under these conditions, the highest symmetry in  the elliptic fibration  is $E_8$ and since 
the  GUT symmetry  in the present work is chosen to be $SU(5)$, 
its commutant is $SU(5)_{\perp}$.  
The physics of the latter is nicely captured by the spectral cover, described by 
a five-degree polynomial
\be 
\label{su5perp}
\begin{split}
{\cal C}_5:&\qquad \sum_{k=0}^5 b_ks^{5-k}=0\,,
\end{split}
\ee
where $b_k$ are holomorphic sections and $s$ is an affine parameter. 
Under the action of certain fluxes and possible monodromies, the polynomial could 
in principle be factorised  to a number of  irreducible components 
\[ {\cal C}_5 \to C_{a_1}\times\cdots \times C_{a_n},\;  1+\cdots +n<5\]
provided that new coefficients preserve the holomorphicity. 
Given the rank of the associated group ($SU(5)_{\perp}$), the simplest possibility is the decomposition into four 
$U(1)$ factors, but this is one among many possibilities. As a matter of fact, in an  F-theory 
context, the roots of the spectral cover equation are related by  non-trivial monodromies.  For 
the $SU(5)_{\perp}$ case at hand, under specific circumstances (related mainly to the properties
of the internal manifold and flux data) these monodromies 
can be described by any possible  subgroup  of the Weyl group $S_5$.  This has
tremendous implications in the effective field theory model,  particularly in
the superpotential couplings. 
The spectral cover equation~(\ref{su5perp})
has roots \(t_i\), which correspond to the weights of \(SU(5)_{\perp}\), i.e. \(b_0\prod_{i=1}^5(s-t_i)=0\). 
The equation describes the matter curves of a particular theory, with roots being related by monodromies depending on the factorisation of this equation.
Thus, we may choose to assume that the spectral cover can be factorised, with new coefficients \(a_j\) that lie within the same field  ${\cal F}$ as \(b_i\). 
Depending on how we factorise, we will see different monodromy groups. 
Motivated by  the peculiar properties of the neutrino sector, here we will attempt to explore the low energy implications of 
the following  factorisations  of the  spectral cover equation 
\be  i)\; {\cal C}_4\times {\cal C}_1,\;\; ii)\; {\cal C}_3\times {\cal C}_2,\;\; iii)\;  {\cal C}_3\times {\cal C}_1\times {\cal C}_1\,.\label{3S4chains}
\ee
Case $i)$ involves the transitive group $S_4$ and its subgroups $A_4 and D_4$ while cases $ii)$ and $iii)$ incorporate the $S_3$, which is isomorphic to $D_3$. For 
later convenience these cases are depicted in figure~\ref{S4chains}.

In case $i)$ for example, the polynomial in equation~(\ref{su5perp}) 
should be separable in the following two  factors 
\begin{equation}
{\cal C}_4\times {\cal C}_1:
\left(a_1+a_2 s+a_3 s^2+a_4 s^3+a_5 s^4\right)(a_6+a_7 s)=0\label{C4C1}
\end{equation}
which implies the  `breaking' of the $SU(5)_{\perp}$ to the monodromy group $S_4$,
(or one of its subgroups such as $A_4$),  described by the fourth degree polynomial 
 \be 
 {\cal C}_4:\qquad  \sum_{k=1}^5 a_ks^{k-1}=0\label{4dp}
 \ee  
 and a \(U(1)\) associated with the linear part. New and old polynomial coefficients satisfy simple relations  $b_k=b_k(a_i)$ which
 can be easily extracted comparing same powers of (\ref{su5perp}) and (\ref{C4C1}) with respect to the parameter $s$.
  Table \ref{table1} summarizes the relations between the coefficients of the unfactorised spectral cover and the \(a_j\) coefficients for 
  the cases under consideration in the present  work.

The homologies of the coefficients \(b_i\) are given in terms of the first Chern class of the tangent bundle (\(c_1\)) and of the normal bundle (\(-t\)),
\begin{equation}\begin{gathered}\left[b_k\right]=\eta-kc_1\,,\\\text{where}\,\,\, \eta=6c_1-t\,.\end{gathered}\end{equation}
We may use these to calculate the homologies $\left[a_j\right]$ of our \(a_j\) coefficients, since if \(b_i=a_j a_k\dots\) then \(\left[b_i\right]=\left[a_j\right]+\left[a_k\right]+\dots\), allowing us to rearrange for the required homologies. Note that since we have in general more \(a_j\) coefficients than our fully determined \(b_i\) coefficients, the homologies of the new coefficients cannot be fully determined.  For example, if we factorise in a \(3+1+1\) arrangement, we must have \(3\) unknown parameters, which we call \(\chi_{k=1,2,3}\).
\begin{table}[t!]
\begin{tabular}{|c|c|c|c|}\hline
\(b_i\) & \(a_j\) coefficients for 4+1&\(a_j\) coefficients for 3+2&\(a_j\) coefficients for 3+1+1\\\hline
\(b_0\)&\(a_5 a_7\)&\(a_4 a_7\)&\(a_4 a_6 a_8\)\\
\(b_1\)&\(a_5 a_6+a_4 a_7\)& \(a_4 a_6 +a_3 a_7\)&\(a_4 a_6 a_7+a_4 a_5a_8+a_3a_6a_8\)\\
\(b_2\)&\(a_4 a_6+a_3 a_7\)&\(a_4 a_5+a_3 a_6+a_2 a_7\)&\(a_4 a_5 a_7+a_3 a_5a_8+a_3a_6a_7+a_2a_6a_8\)\\
\(b_3\)&\(a_3 a_6+a_2 a_7\)&\(a_3 a_5+a_2 a_6+a_1 a_7\)&\(a_3 a_5 a_7+a_2 a_5a_8+a_2a_6a_7+a_1a_6a_8\) \\
\(b_4\)&\(a_2 a_6 +a_1 a_7\)&\(a_2 a_5+ a_1 a_6\)&\(a_2a_5a_7+a_1a_6a_7+a_1a_5a_8\) \\
\(b_5\)&\(a_1 a_6\)&\(a_1 a_5\)&\(a_1a_5a_7\) \\
\hline
\end{tabular}\caption{table showing the relations $b_i=b_i(a_j)$ between coefficients of the spectral cover equation under various decompositions from the unfactorised equation.\label{table1}}\end{table}
In the following sections we will examine in detail the predictions of the $A_4$ and $S_3$ models.

\section{$A_4$  models in F-theory}

We assume that the spectral cover equation factorises to a quartic polynomial and a linear part, as shown in~(\ref{C4C1}). T
he homologies of the new coefficients may be derived from the original \(b_i\) coefficients. Referring to Table \ref{table1}, we can see that the homologies for this factorisation are easily calculable, up to some arbitrariness of one of the coefficients - we have seven \(a_j\) and only six \(b_i\). We choose \([a_6]=\chi\) in order to make this tractable. It can then be shown that the homologies obey:
\begin{equation}\begin{gathered}\left[a_i\right]=\eta-(6-i)c_1-\chi\,,\\\text{Where }i\in\{1,\dots,5\}\\
\left[a_7\right]=c_1+\chi\\
\left[a_6\right]=\chi\,.
\end{gathered}\end{equation}
This amounts to asserting that the five of \(SU(5)_{\perp}\) `breaks' to a discrete symmetry between four of its weights (\(S_4\) or one of its subgroups) and a \(U(1)_{\perp}\).\\

The roots of the spectral cover equation must obey:
\begin{equation}\label{tens}b_0\prod_{i=1}^{5}(s-t_i)=0\,,\end{equation}
where \(t_i\) are the weights of the five representation of \(SU(5)_{\perp}\). When \(s=0\), this defines the tenplet matter curves of the  \(SU(5)_{\text{GUT}}\) \cite{Donagi:2009ra}, with the number of curves being determined by how the result factorises. In the case under consideration, when \(s=0\), \(b_5=0\). After referring to Table \ref{table1}, we see that this implies that \(P_{10}=a_1 a_6=0\). Therefore there are two tenplet matter curves, whose homologies are given by those of \(a_1\) and \(a_6\). We shall assume at this point that these are the only two distinct curves, though \(a_1\) appears to be associated with \(S_4\) (or a subgroup) and hence should be reducible to a triplet and singlet.

Similarly, for the fiveplets, we have 
\begin{equation}\label{fives}b_0\prod_{i=1}^{5}(s-t_i-t_j)=0\,\,\,\,\text{for}\,i\neq j,\end{equation}
which can be shown\footnote{See for example~\cite{Donagi:2009ra}.} to give the defining condition for the fiveplets: \(P_5=b_4b_3^2-b_2b_5b_3+b_0b_5^2=0\). Again consulting the  Table~\ref{table1}, we can write this in terms of the \(a_j\) coefficients: \begin{equation}P_5=\left(a_3 a_6+a_2 a_7\right)^2\left(a_2 a_6 a_1 a_7\right)-\left(a_4 a_6+a_3 a_7\right)\left(a_3 a_6+a_2 a_7\right)\left(a_1a_6\right)+a_1^2a_5a_6^2a_7\,.\end{equation}
Using the condition that \(SU(5)\) must be traceless, and hence \(b_1=0\),  we have that \(a_4a_7+a_5a_6=0\). An Ansatz solution of this condition is \(a_4=\pm a_0a_6\) and \(a_5=\mp a_0a_7\), where \(a_0\) is some appropriate scaling with homology \(\left[a_0\right]=\eta-2(c_1+\chi)\), which is trivially derived from the homologies of \(a_4\) and \(a_6\) (or indeed \(a_5\) and \(a_7\)) \cite{Antoniadis:2013joa}. If we introduce this, then \(P_5\) splits into two matter curves:
\begin{equation}P_5=\left(a_2^2a_7+a_2a_3a_6\mp a_0a_1a_6^2\right)\left(a_3 a_6^2+(a_2a_6+a_1a_7)a_7\right)=0\,.\end{equation}
The homologies of these curves are calculated from those of the \(b_i\) coefficients and are presented in Table \ref{table2}. We may also impose flux restrictions if we define:
\begin{equation}\begin{gathered}
{\mathcal F}_Y\cdot\chi=N\,,\\
{\mathcal F}_Y\cdot c_1= 
{\mathcal F}_Y\cdot\eta=0\,,
\end{gathered}
\end{equation}
where \(N\in\mathbb{Z}\) and \(\mathcal{F}_Y\) is the hypercharge flux. 
\\
\begin{table}[t!]\centering
\begin{tabular}{|l|c|c|c|c|}\hline
Curve & Equation & Homology&Hyperflux - N &Multiplicity\\\hline
\(10_a\)&\(a_1\) & \(\eta -5c_1-\chi\)&\(-N\)&\(M_{10_a}\)\\
\(10_b\)&\(a_6\) & \(\chi\)& \(+N\)&\(M_{10_b}\) \\
\(5_c\)&\(a_2^2a_7+a_2a_3a_6\mp a_0a_1a_6^2\)&\(2\eta-7c_1-\chi\)&\(-N\)&\(M_{5_c}\)\\
\(5_d\)&\(a_3 a_6^2+(a_2a_6+a_1a_7)a_7\)&\(\eta-3c_1+\chi\)&\(+N\)&\(M_{5_d}\)\\
\hline
\end{tabular}\caption{table of matter curves, their homologies, charges and multiplicities.\label{table2}}\end{table}

Considering equation \eqref{tens}, we see that \(b_5/b_0=t_1t_2t_3t_4t_5\), so there are at most five ten-curves, one for each of the weights. Under \(S_4\) and it's subgroups, four of these are identified, which corroborates with the two matter curves seen in Table \ref{table1}. As such we identify \(t_{i=1,2,3,4}\) with this monodromy group and the coefficient \(a_1\) and leave \(t_5\) to be associated to \(a_6\). \\

Similarly, equation \eqref{fives} shows that we have at most ten five-curves when \(s=0\), given in the form \(t_i+t_j\) with \(i\neq j\). Examining the equations for the two five curves that are manifest in this model after application of our monodromy,  the quadruplet involving \(t_i+t_5\) forms the curve labeled \(5_d\), while the remaining  sextet - $t_i+t_j$ with $i,j\neq5$ - sits on the \(5_c\) curve.

\subsection{The discriminant }

The above  considerations  apply equally to both the $S_4$ as well as $A_4$ discrete groups.  
 From the effective model point of view, all the
  useful information  is  encoded in the properties of the polynomial coefficients $a_k$
 and if we wish to distinguish these two models further assumptions for the latter
 coefficients have to be made. Indeed, if we assume that in the above polynomial, 
the  coefficients belong to a certain field $a_k\in {\cal F}$, 
without imposing any additional specific  restrictions on $a_k$, the roots  
  exhibit an $S_4$ symmetry.  If, as desired, the symmetry acting on  roots is 
 the  subgroup  $A_4$ the coefficients  $a_k$ must respect certain conditions. 
 Such constraints emerge from the study of partially symmetric functions of roots. In 
 the present case in particular, we recall that the $A_4$ discrete symmetry is associated 
 only to even permutations  of the four roots $t_i$. Further, we note now  that
 the partially symmetic function 
 \[ \delta = (t_1-t_2) (t_1-t_3) (t_1-t_4) (t_2-t_3) (t_2-t_4) (t_3-t_4)\]
 is invariant only under the even permutations of roots. The quantity $\delta$  is the 
 square root of the discriminant, 
 \be 
 \Delta =\delta^2
 \label{Dd2}
 \ee
  and as such $\delta$
 should be written as a function of the polynomial coefficients $a_k\in {\cal F}$ 
 so that $\delta \in {\cal F}$ too.  The discriminant is computed by standard formulae
 and is found to be
 \be
 \label{Disc}
 \begin{split}
  \Delta(a_k)&=
  256 a_1^3 a_5^3-\left(27 a_2^4-144 a_1 a_3 a_2^2+192 a_1^2 a_4 a_2+128 a_1^2 a_3^2\right) a_5^2\\
  &-2 \left(2 \left(a_2^2-4 a_1 a_3\right) a_3^3- \left(9 a_2^2-40 a_1 a_3\right)a_2 a_4 a_3+3 \left(a_2^2-24 a_1 a_3\right)a_1  a_4^2\right) a_5\\
  &-a_4^2 \left(4 a_4 a_2^3+a_3^2
     a_2^2-18 a_1 a_3 a_4 a_2+\left(4 a_3^3+27 a_1 a_4^2\right)a_1 \right)
  \end{split}
  \ee
  In order to examine the implications  of  (\ref{Dd2})
 we write the discriminant as a polynomial of the coefficient $a_3$~\cite{Antoniadis:2013joa}
 \be
 \Delta \equiv g(a_3)= \sum_{n=0}^4 c_n a_3^n
 \ee
 where the  $c_n$ are functions of the  remaining coefficients $a_k,\; k\ne 3$ and
 can be easily computed by comparison with (\ref{Disc}).
 We may equivalently demand that $ g(a_3)$ is a square of a second degree
  polynomial 
  \[g(a_3) =(\kappa a_3^2+\lambda a_3+\mu)^2\]
 A necessary condition that the polynomial $g(a_3)$ is a square, is its own discriminant 
$\Delta_{g}$ to be  zero.  One finds
\[ \Delta_{g}\propto D_1^2D_2^3\]  where
 \be
 \label{Df12}
 \begin{split}
 D_1&=a_2^2a_5-a_1a_4^2\\
 D_2&=\left(27a_1^2 a_4- a_2^3 \right)a_4^3-6 a_1 a_2^2 a_5
 a_4^2+3  a_2 \left(9 a_2^3-256 a_1^2 a_4\right) a_5^2+4096 a_1^3 a_5^3
    \end{split}
    \ee
    We observe that there are two ways to eliminate the discriminant of the
    polynomial, either putting $D_1=0$ or by demanding $D_2=0$~\cite{Antoniadis:2013joa}.

 In the first case, we can achieve $\Delta=\delta^2$ if we solve the constraint
 $D_1=0$ as follows
\be 
\label{D1sol}
\begin{split}
a_2^2&=2a_1a_3\\
a_4^2&=2a_3a_5
\end{split}
\ee
 Substituting  the solutions (\ref{D1sol})  in the discriminant we  find 
\be
 \Delta \;= \;\delta^2\;=\;\left[a_2 a_4 \left(a_3^2-2\, a_2 a_4\right) \left(a_3^2-a_2 a_4\right)/a_3^3\right]^2
\ee
The above  constitute the necessary conditions to obtain the  reduction of the symmetry~\cite{Antoniadis:2013joa} down to the Klein group
$V\sim Z_2\times Z_2$. 
On the other hand, the second condition   $D_2=0$, implies a non-trivial relation among the coefficients
\ba
( a_2^2a_5-a_4^2a_1)^2&=&\left(\frac{a_2a_4-16a_1a_5}{3}\right)^3\label{cubic}
\ea
 Plugging in the $b_1=0$ solution,   the constraint (\ref{cubic})
take the form
\ba
( a_2^2a_7+a_0a_1a_6^2)^2&=&a_0\left(\frac{a_2a_6+16a_1a_7}{3}\right)^3\label{cubic1}
\ea
which is just the condition on the polynomial coefficients to obtain the transition $S_4\to A_4$.

\subsection{Towards an $SU(5)\times A_4$ model }

Using the previous analysis, in this section we will present a specific example based on the $SU(5)\times A_4\times U(1)$ symmetry. 
We will make specific choices of the flux parameters and derive the spectrum and its superpotential, focusing in particular on the neutrino sector.\\

It can be shown that if we assume an \(A_4\) monodromy any quadruplet is reducible to a triplet and singlet representation, while the sextet of the fives reduces to two triplets (details can be found in the
appendix). 
\subsubsection{Singlet-Triplet Splitting Mechanism}

It is known from group theory and a physical understanding of the group that the four roots forming the basis under \(A_4\) may be reduced to a singlet and triplet. As such we might suppose intuitively that the quartic curve of \(A_4\) decomposes  into two curves - a singlet and a triplet of \(A_4\). \\

As a mechanism for this we consider an analogy to the breaking of the \(SU(5)_{GUT} \) group by \(U(1)_Y\). We then postulate a mechanism to facilitate Singlet-Triplet splitting in a similar vein. Switching on a flux in some direction of the perpendicular group, we propose that the singlet and triplet of \(A_4\) will split to form two curves. This flux should be proportional to one of the generators of \(A_4\), so that the broken group commutes with it. If we choose to switch on \(U(1)_s\) flux in the direction of the singlet of \(A_4\), then the discrete symmetry  will remain unbroken by this choice. \\

Continuing our previous analogy, this would split the curve as follows:
\begin{equation}
(10,4)=\left\{\begin{gathered}(10,1) = M+N_s\\(10,3)=M\end{gathered}\right.\,.
\end{equation}•

The homologies of the new curves are not immediately known. However, they can be constrained by the previously known homologies given in Table \ref{table2}. The coefficient describing the curve should be expressed as the product of two coefficients, one describing each of the new curves - \(a_i=c_1c_2\). As such, the homologies of the new curves will be determined by \([a_i]=[c_1]+[c_2]\). \\

If we assign the \(U(1)\) flux parameters by hand, we can set the constraints on the homologies of our new curves. For example, for the curve given in Table \ref{table2} as \(10_a\) would decompose into two curves - \(10_1\) and \(10_2\), say. Assigning the flux parameter, \(N\), to the \(10_2\) curve, we constrain the homologies of the two new curves as follows:
\begin{align*}
[10_1]=&a\eta+bc_1\\
[10_2]=&c\eta+dc_1-\chi\\
\text{Where: }a+c=1\,&\text{ and }\,b+d=-5\,.
\end{align*}
Similar constraints may also be placed on the five-curves after decomposition.\\

Using our procedure, we can postulate that the charge \(N\) will be associated to the singlet curve by the mechanism of a flux in the singlet direction. This protects the overall charge of \(N\) in the theory. With the fiveplet curves it is not immediately clear how to apply this since the sextet of \(A_4\) can be shown to factorise into two triplets. Closer  examination points to the necessity to cancel anomalies. As such the curves carrying \(H_u\) and \(H_d\) must both have the same charge under \(N\). This will insure that they cancel anomalies correctly. These motivating ideas have been applied in Table \ref{table3}.

\subsubsection{GUT-group doublet-triplet splitting}
 Initially massless states residing on the matter curves comprise complete vector multiplets. Chirality is generated by switching on appropriate fluxes. At the $SU(5)$ level, we assume the existence of $M_{5}$  fiveplets and $M_{10}$  tenplets.
The multiplicities are not entirely independent, since we require anomaly cancellation,\footnote{For a discussion in relaxing 
 some of the anomaly cancellation conditions and related issues see~\cite{Mayrhofer:2013ara}.} which amounts to the requirement that \(\sum_i M_{5_i}+\sum_j M_{10_j}=0\).
Next, turning on the hypercharge  flux, under the $SU(5)$ symmetry breaking the $10$ and $5,\bar 5$
 representations split into different numbers of Standard Model multiplets \cite{Marsano:2010sq}.
Assuming  $N$ units of hyperflux piercing a given matter curve,  the fiveplets  split according to:
\begin{equation}\begin{gathered}n(3,1)_{-1/3}-n(\bar{3},1)_{+1/3}=M_5\,,\\
 n(1,2)_{+1/2}-n(1,2)_{-1/2}=M_5+N\,,\end{gathered}\end{equation}
Similarly, the  $M_{10}$  tenplets decompose under the influence of $N$ hyperflux units to the
  following SM-representations:
\begin{equation}\begin{gathered}n(3,2)_{+1/6}-n(\bar{3},2)_{-1/6}=M_{10}\,,\\
 n(\bar{3},1)_{-2/3}-n(3,1)_{+2/3}=M_{10}-N\,,\\
n(1,1)_{+1}-n(1,1)_{-1}=M_{10}+N\,.\end{gathered}\end{equation}
 
 Using the relations for the multiplicities of our matter states, we can construct a model with the spectrum
parametrised in terms of a few integers in a manner presented in Table \ref{table3}. \\
\begin{table}[t]\centering\begin{tabular}{|c|c|c|c|c|c|}\hline
Curve& \(SU(5) \times A_4\times U(1)_{\perp}\)  & \(N_Y\) &M & Matter content& R\\\hline
\(10_1\) & \((10,3)_0\) & \(0\) & \(M_{T1}\) & \(3\left[M_{T1}Q_L+u^c_L(M_{T1}-N_Y)+e^c_L(M_{T1}+N_Y)\right]\)&\(1\)\\
\(10_2\) & \((10,1)_0\) &\(-N\) & \(M_{T2}\) & \(M_{T2}Q_L+u^c_L(M_{T2}-N_Y)+e^c_L(M_{T2}+N_Y)\)&\(1\)\\
\(10_3\) & \((10,1)_{t_5}\) & \(+N\) & \(M_{T3}\) & \(M_{T3}Q_L+u^c_L(M_{T3}-N_Y)+e^c_L(M_{T3}+N_Y)\)&\(1\)\\
\(5_1\) & \((5,3)_0\) & \(0\) & \(M_{F1}\) & \(3\left[M_{F1}\bar{d}^c_L+(M_{F1}+N_Y)\bar{L}\right]\) &\(1\)\\
\(5_2\) & \((5,3)_0\) & \(-N\) & \(M_{F2}\) & \(3\left[M_{F2}\bar{\bar{D}}+(M_{F2}+N_Y)\bar{H}_d)\right]\) &\(0\)\\
\(5_3\) & \((5,3)_{t_5}\) & \(+N\) & \(M_{F3}\) &  \(3\left[M_{F3}D+(M_{F3}+N_Y)H_u\right]\) &\(0\)\\
\(5_4\) & \((5,1)_{t_5}\) & \(0\) & \(M_{F4}\) & \(M_{F4}\bar{d}^c_L+(M_{F4}+N_Y)\bar{L}\) &\(1\)\\
\hline
\end{tabular}\caption{Table showing the possible matter content for an \(SU(5)_{\text{GUT}}\times A_4\times U(1)_{\perp}\), where it is assumed the reducible representation of the monodromy group may split the matter curves. The curves are also assumed to have an R-symmetry\label{table3}}
\end{table}

In order to curtail the number of possible couplings and suppress operators surplus to requirement, we also call on the services of an R-symmetry. This is commonly found in supersymmetric models, and requires that all couplings have a total R-symmetry of 2. Curves  carrying SM-like fermions are taken to have \(R=1\), with all other curves \(R=0\).
%
%
%
%

\subsection{A simple model: \({N=0}\)}
Any realistic model based on this table must contain at least 3 generations of quark matter (\(10_{M_i}\)), 3 generations of leptonic matter (\(\bar{5}_{M_i}\)), and one each of \(5_{H_u}\) and \(5_{H_d}\). We shall attempt to construct a model with these properties using simple choices for our free variables. 

In order to build a simple model, let us first choose the simple case where N=0, then we make the following assignments:
\begin{equation}\begin{gathered}M_{T1}=M_{F4}=0\\ M_{T2}=1\\M_{T3}=2\\M_{F1}=M_{F2}=-M_{F3}=-1\end{gathered}\end{equation}
\begin{table}[t!]\centering\begin{tabular}{|c|c|c|c|c|}\hline
Curve&\(SU(5)\times A_4\times U(1)\)&M&Matter content&R-Symmetry\\\hline
\(10_1\) & \((10,3)_0\)&0 & -&1 \\
\(10_2=T_3\) & \((10,1)_0\) &1&\(Q_L +u^c_L+e^c_L\)&1\\
\(10_3=T\) & \((10,1)_{t_5}\) &2&\(2Q_L +2u^c_L+2e^c_L\)&1\\
\(\bar{5}_1=F\) & \((\bar{5},3)_0\) &1&\(3L + 3d^c_L\)&1\\
\(\bar{5}_2=H_d\) & \((\bar{5},3)_0\) &1&\(3\bar{D}+3H_d\)&0\\
\(5_3=H_u\) & \((5,3)_{t_5}\) &1&\(3D+3H_u\)&0\\
\(5_4\) & \((5,1)_{t_5}\) &0&  -&1 \\
\(\theta_a\)&\((1,3)_{-t_5}\)&-&  Flavons&0 \\
\(\theta_b\)&\((1,1)_{-t_5}\)&-& Flavon&0 \\
\(\theta_c\)&\((1,3)_0\)&-& \(\nu_R\) &1\\
\(\theta_d\)&\((1,3)_0\)&-& Flavons&0 \\
\(\theta_{a'}\)&\((1,3)_{t_5}\)&-& -&0 \\
\(\theta_{b'}\)&\((1,1)_{t_5}\)&-& -&0 \\\hline
\end{tabular}\caption{Table of Matter content in \(N=0\) model\label{table4}}\end{table}

Note that it does not immediately appear possible to select a matter arrangement that provides a renormalisable top-coupling, since we will be required to use our GUT-singlets to cancel residual \(t_5\) charges in our couplings,  at the cost of renormalisability.

\subsection{Basis}
The bases of the triplets are such that triplet products, \(3_a\times3_b =1+1'+1''+3_1+3_2\),  behave as:
\[\begin{gathered}
1=a_1b_2+a_2b_2+a_3b_3\\
1'=a_1b_2+\omega a_2b_2+\omega^2a_3b_3\\
1''=a_1b_2+\omega^2a_2b_2+\omega a_3b_3\\
3_1=(a_2b_3,\,a_3b_1,\,a_1b_2)^{\text{T}}\\
3_2=(a_3b_2,\,a_1b_3,\,a_2b_1)^{\text{T}}\end{gathered}\]
where \(3_a=(a_1,\,a_2,\,a_3)^{\text{T}}\) and \(3_b=(b_1,\,b_2,\,b_3)^{\text{T}}\). This has been demonstrated in the Appendix \ref{a4basis}, where we show that the quadruplet of weights decomposes to a singlet and triplet in this basis. Note that all couplings must of course produce singlets of \(A_4\) by use of these triplet products where appropriate.

\begin{table}[t!]\centering\begin{tabular}{|c|c | c|}\hline
Coupling type &Generations & Full coupling\\\hline
Top-type&Third generation& \(T_3\cdot T_3\cdot H_u\cdot\theta_a\)\\

&Third-First/Second generation&\(T\cdot T_3\cdot H_u\cdot\theta_a\cdot\theta_b\)\\
&&\(T\cdot T_3\cdot H_u\cdot(\theta_a)^2\)\\

&First/Second generation&\(T\cdot T\cdot H_u\cdot\theta_a\cdot(\theta_b)^2\)\\
&&\( T\cdot T\cdot H_u\cdot(\theta_a)^2\cdot\theta_b\)\\
&&\(T\cdot T\cdot H_u\cdot(\theta_a)^3\)\\

\hline

Bottom-type/Charged Leptons&Third generation& \(F\cdot H_d\cdot T_3\)\\
&&\(F\cdot H_d\cdot T_3 \cdot \theta_d\)\\
&First/Second generation&\(F\cdot H_d\cdot T\cdot\theta_b\)\\
&& \(F\cdot H_d\cdot T\cdot\theta_a\)\\
&& \(F\cdot H_d\cdot T\cdot\theta_a\cdot\theta_d\)\\
&& \(F\cdot H_d\cdot T\cdot\theta_b\cdot\theta_d\)\\
\hline

Neutrinos& Dirac-type mass&\(\theta_c\cdot F\cdot H_u\cdot\theta_a\)\\
&&\(\theta_c\cdot F\cdot H_u\cdot\theta_a\cdot\theta_d\)\\
&&\(\theta_c\cdot F\cdot H_u\cdot\theta_b\)\\
&&\(\theta_c\cdot F\cdot H_u\cdot\theta_b\cdot\theta_d\)\\
&Right-handed neutrinos&\(M\theta_c\cdot\theta_c\)\\
&&\((\theta_d)^n\cdot\theta_c\cdot\theta_c\)\\\hline
\end{tabular}\caption{Table of all mass operators for \(N=0\) model. 
 \label{table5}}\end{table}

\subsection{Top-type quarks}
The Top-type quarks admit a total of six mass terms, as shown in Table \ref{table5}. 
The third generation has only one valid Yukawa coupling -  \(T_3\cdot T_3\cdot H_u\cdot\theta_a\). Using the above algebra, we find that this coupling is:
\begin{align*}(1\times1)\times(3\times3)&\rightarrow 1\times 1\\&\rightarrow1\\
(T_3\times T_3)\times H_u\times\theta_a&\rightarrow (T_3\times T_3) v_ia_i \\
&i=1,2,3
\end{align*}

With the choice of vacuum expectation values (VEVs):
\begin{equation}\begin{gathered}\langle H_u\rangle=(v,0,0)^{\text{T}}\\\langle\theta_a\rangle=(a,0,0)^{\text{T}}\\\langle \theta_b\rangle=b\end{gathered}\end{equation}
this will give the Top quark it's mass, \(m_t=yva\). The choice is partly motivated by $A_4$ algebra, as the VEV will preserve the S-generators. This choice of VEVs will also kill off the the operators \(T\cdot T_3\cdot H_u\cdot(\theta_a)^2\) and \( T\cdot T\cdot H_u\cdot(\theta_a)^2\cdot\theta_b\), which can be seen by applying the algebra above. \\

The full algebra of the contributions from the remaining operators is included in Appendix \ref{ykwalgebra}. Under the already assigned VEVs, the remaining operators contribute to give the overall mass matrix for the Top-type quarks:
\begin{equation}m_{u,c,t}= va\left(\begin{array}
{ccc}
y_3b^2+y_4a^2&y_3b^2+y_4a^2&y_2b\\
y_3b^2+y_4a^2&y_3b^2+y_4a^2&y_2b\\
y_2b&y_2b&y_1
\end{array}\right)\end{equation}
This matrix is clearly hierarchical with the third generation dominating the hierarchy, since the couplings should be suppressed by the higher order nature of the operators involved.  Due to the rank theorem \cite{Cecotti:2009zf}, the two lighter generations can only have one massive eigenvalue. However, corrections due to instantons and non-commutative fluxes are known as mechanisms to recover a light mass for the first generation \cite{Cecotti:2009zf}\cite{Aparicio:2011jx}.

\subsection{Charged Leptons}
The Charged Lepton and Bottom-type quark masses come from the same GUT operators. Unlike the Top-type quarks, these masses will involve SM-fermionic matter that lives on curves that are triplets under \(A_4\). It will be possible to avoid unwanted relations between these generations using the ten-curves, which are strictly singlets of the monodromy group. The operators, as per Table \ref{table5}, are computed in full in Appendix \ref{ykwalgebra}.\\

Since we wish to have a reasonably hierarchical structure, we shall require that the dominating terms be in the third generation. This is best served by selecting the VEV \(\langle H_d\rangle=(0,0,v)^{\text{T}}\). Taking the lowest order of operator to dominate each element, since we have non-renormalisable operators, we see that we have then: 
\begin{equation}m_{e,\mu,\tau}=v\left(\begin{array}
{ccc}y_7d_2b+y_{11}d_3a&y_7d_2b+y_{11}d_3a&y_3d_2\\
y_5a&y_5a&y_2d_1\\
y_4b&y_4b&y_1
\end{array}\right)\,.
\end{equation}

We should again be able to use the Rank Theorem to argue that while the first generation should not get a mass by this mechanism, the mass may be generated by other effects  \cite{Cecotti:2009zf}\cite{Aparicio:2011jx}. We also expect there might be small corrections due to the higher order contributions, though we shall not consider these here. \\

The bottom-type quarks in $SU(5)$ have the same masses as the
charged leptons, with the exact relation between the Yukawa matrices being due to a transpose.
However   this fact is known to be inconsistent with experiment.
    In general,  when    renormalization group running effects are
taken into account,
    the problem can be evaded only for the third generation. Indeed,
     the mass relation $m_b = m_{\tau}$ at $M_{GUT}$  can be
    made consistent with the low energy measured ratio $m_b/m_{\tau}$
    for suitable values of $\tan\beta$.  In field theory $SU(5)$ GUTs
    the  successful Georgi-Jarlskog GUT  relation $m_s/m_{\mu}=1/3$
    can be obtained from  a term involving the representations $\bar
5\cdot 10\cdot 45$ but
    in the F-theory context this is not possible due to the absence of the
    45 representation. Nevertheless, the order one Yukawa coefficients
may be
    different because the intersection points need not be at the
    same enhanced symmetry point. The final structure of the mass matrices
    is revealed   when flux and other threshold effects are taken into
account.
    These issues will not be discussed further here and a more detailed
exposition may be found in \cite{Leontaris:2010zd}, with other useful discussion to be found in \cite{Font:2012wq}.

\subsection{Neutrino sector}
Neutrinos are unique in the realms of currently known matter in that they may have both Dirac and Majorana mass terms. The couplings for these must involve an \(SU(5)\) singlet to account for the required right-handed neutrinos, which we might suppose is \(\theta_c=(1,3)_0\). It is evident from Table \ref{table5} that the Dirac mass is the formed of a handful of couplings at different orders in operators. We also have a Majorana operator for the right-handed neutrinos, which will be subject to corrections due to the \(\theta_d\) singlet, which we assign the most general VEV, \(\langle\theta_d\rangle=(d_1,d_2,d_3)^{\text{T}}\). 

If we now analyze the operators for the neutrino sector in brief, the two leading order contribution are from the \(\theta_c\cdot F\cdot H_u\cdot \theta_a\) and \(\theta_c\cdot F\cdot H_u\cdot \theta_b\) operators. With the VEV alignments \(\langle\theta_a\rangle=(a,0,0)^{\text{T}}\) and \(\langle H_u\rangle=(v,0,0)^{\text{T}}\), we have a total matrix for these contributions that displays strong mixing between the second and third generations: 
\begin{equation}
m=\left(\begin{array}{ccc}
y_0va&0&0\\
0&y_1va&y_9bv\\
0&y_8bv&y_1va
\end{array}\right)\,,
\end{equation}
where \(y_0=y_1+y_2+y_3\). The higher order operators, \(\theta_c\cdot F\cdot H_u\cdot\theta_a\cdot\theta_d\) and \(\theta_c\cdot F\cdot H_u\cdot\theta_b\cdot\theta_d\), will serve to add corrections to this matrix, which may be necessary to generate mixing outside the already evident large 2-3 mixing from the lowest order operators. If we consider the  \(\theta_c\cdot F\cdot H_u\cdot\theta_b\cdot\theta_d\) operator, 
\begin{equation}
\theta_c\cdot F\cdot H_u\cdot\theta_d\cdot\theta_b
\rightarrow \left(\begin{array}{ccc}0&z_3vd_2b&z_2vd_3b\\z_1vd_2b&0&0\\z_4vd_3b&0&0\end{array}\right)
\end{equation}
We use \(z_i\) coefficients to denote the suppression expected to affect these couplings due to renormalisability requirements. We need only concern ourselves with the combinations that add contributions to the off-diagonal elements where the lower order operators have not given a contribution, as these lower orders should dominate the corrections. Hence, the remaining allowed combinations will not be considered for the sake of simplicity. If we do this we are left a matrix of the form:
\begin{equation}\label{dm}M_D=\left(
\begin{array}
{ccc}y_0va&z_3vd_2b&z_2vd_3b\\
z_1vd_2b&y_1va&y_9bv\\
z_4vd_3b&y_8bv&y_1va
\end{array}\right)
\end{equation} \\

The right-handed neutrinos admit Majorana operators of the type \(\theta_c\cdot\theta_c\cdot(\theta_d)^n\), with \(n\in\{0,1,\dots\}\). The \(n=0\) operator will fill out the diagonal of the mass matrix, while the \(n=1\) operator fills the off-diagonal. Higher order operators can again be taken as dominated by these first two, lower order operators. The Majorana mass matrix can then be used along with the Dirac mass matrix in order to generate light effective neutrino masses via a see-saw mechanism.

\begin{equation}
M_R=M\left(\begin{array}
{ccc}1&0&0\\0&1&0\\0&0&1
\end{array}\right)+ y\left(\begin{array}{ccc}0&d_3&d_2\\ d_3&0&d_1\\ d_2&d_1&0
\end{array}\right)
\end{equation}• \\

The Dirac mass matrix can be summarised as in equation \eqref{dm}. This matrix is rank 3, with a clear large mixing between two generations that we expect to generate a large \(\theta_{23}\). In order to reduce the parameters involved in the effective mass matrix, we will simplify the problem by searching only for solutions where \(z_1=z_3\) and \(z_2=z_4\), which significantly narrows the parameter space. We will then define some dimensionless parameters that will simplify the matrix:
\begin{align}
Y_1=&\frac{y_1}{y_0}\le1\\
Y_{2,3}=&\frac{y_{8,9}b}{y_0a}\\
Z_{1}=&\frac{z_1d_2b}{y_0a}\\
Z_2=&\frac{z_2d_3b}{y_0a}
\end{align}
If we implement these definitions, we find the Dirac mass matrix becomes:
\begin{equation}M_D=y_0va\left(
\begin{array}{ccc}
1&Z_1&Z_2\\
Z_1&Y_1&Y_3\\
Z_2&Y_2&Y_1
\end{array}\right)
\end{equation}
\begin{table}[t!]\centering
\begin{tabular}{|c|c|c|}\hline
&Central value&Min \(\rightarrow\) Max\\\hline
\(\theta_{12}/^{\circ}\)&33.57&32.82\(\rightarrow\)34.34\\
\(\theta_{23}/^{\circ}\)&41.9&41.5\(\rightarrow\)42.4\\
\(\theta_{13}/^{\circ}\)&8.73&8.37\(\rightarrow\)9.08\\
\(\Delta m_{21}^2/10^{-5}\text{eV}\)&\(7.45\)&\(7.29\rightarrow7.64\)\\
\(\Delta m_{31}^2/ 10^{-3}\text{eV}\)&\(2.417\)&\(2.403\rightarrow2.431\)\\
\(R=\frac{\Delta m_{31}^2}{\Delta m_{21}^2}\)&32.0&\(31.1\rightarrow33.0\)\\
\hline
\end{tabular}\caption{Summary of neutrino parameters, using best fit values as found at nu-fit.org, the work of which relies upon \cite{GonzalezGarcia:2012sz} \label{nures}.}
\end{table}
The Right-handed neutrino Majorana mass matrix can be approximated if we take only the \(\theta_c\cdot\theta_c\) operator, since this should give a large mass scale to the right-handed neutrinos and dominate the matrix. This will leave the Weinberg operator for effective neutrino mass,  \(M_{eff}=M_DM_R^{-1}M_D^{\text{T}}\), as:
\begin{align}
M_{eff}=m_0&\left(\begin{array}{ccc}
1+Z_1^2+Z_2^2&Y_1Z_1+Y_3Z_2+Z_1&Y_2Z_1+Y_1Z_2+Z_2\\
Y_1Z_1+Y_3Z_2+Z_1&Y_1^2+Y_3^2+Z_1^2&Y_1(Y_2+Y_3)+Z_1Z_2\\
Y_2Z_1+Y_1Z_2+Z_2&Y_1(Y_2+Y_3)+Z_1Z_2&Y_1^2+Y_2^2+Z_2^2
\end{array}\right)\,,
\end{align}
Where we have also defined a mass parameter:
\begin{equation}
m_0=\frac{y_0^2v^2a^2}{M}\,,
\end{equation}•

We then proceed to diagonalise this matrix computationally in terms of three mixing angles as is the standard procedure \cite{King:2003jb}, before attempting to fit the result to experimental inputs. \\
\subsection{Analysis}
We shall focus on the ratio of the mass squared differences:
\begin{equation}
R=\left|\frac{m_3^2-m_2^2}{m_2^2-m_1^2}\right|\,,
\end{equation}
which is known due to the well measured mass differences, \(\Delta m_{32}^2\) and \(\Delta m_{21}^2\) \cite{GonzalezGarcia:2012sz}. These give us a value of \(R\approx 32\), which we may solve for numerically in our model using Mathematica or another suitable maths package. If we then fit the optimised values to the mass scales measured by experiment, we may predict absolute neutrino masses and further compare them with cosmological constraints.\\

\begin{figure}[t!]\centering
\includegraphics[scale=0.55]{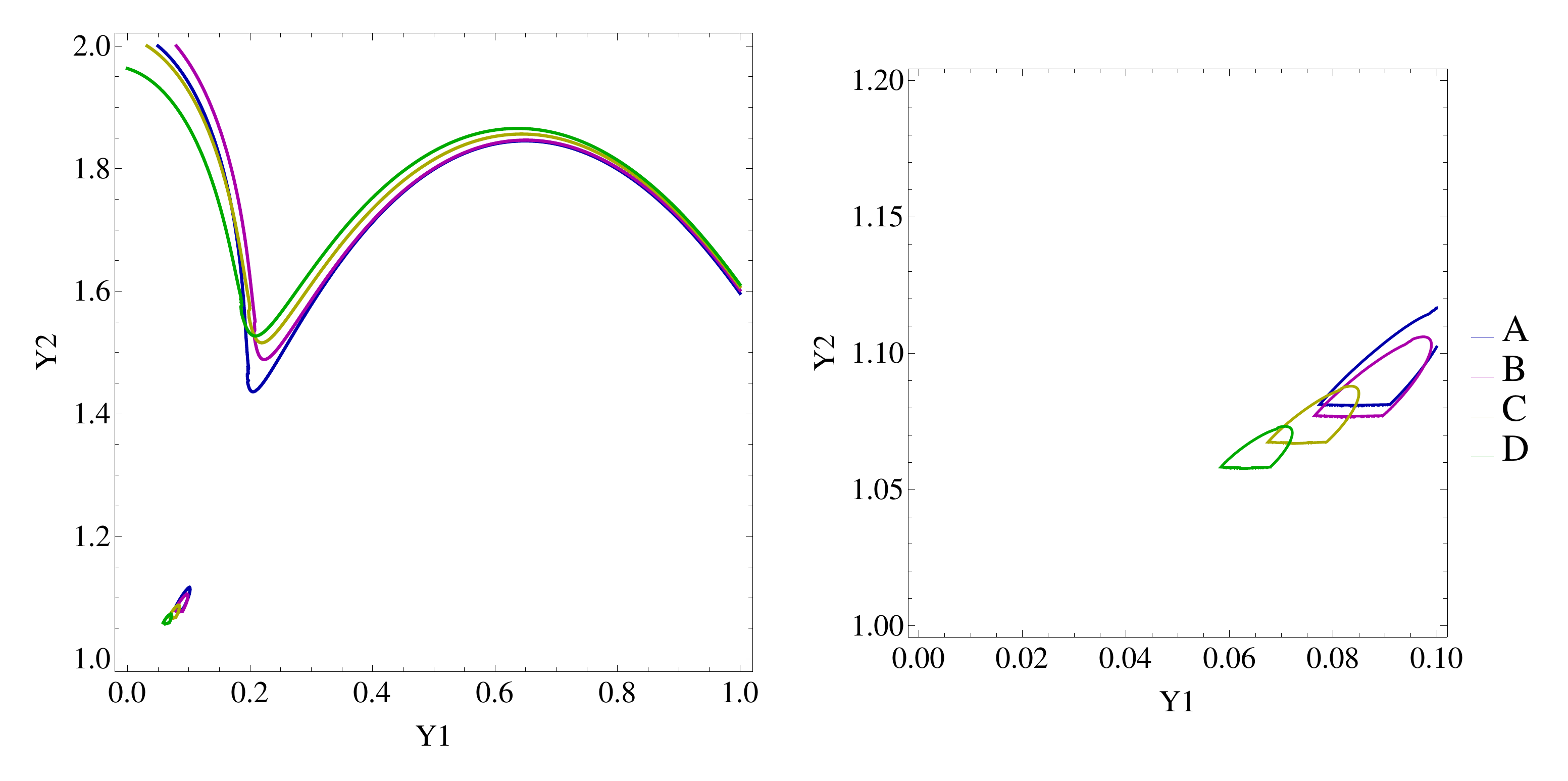}
\caption{Plots of lines with the best fit value of \(R=32\) in the parameter space of \((Y_1,Y_2)\). Left: The full range of the space examined. Right: A close plot of a small portion of the parameter space taken from the full plot. The curves have \((Y_3, Z_1, Z_2)\) values set as follows: \(A=(1.08, 0.05, 0.02)\), \(B=(1.08, 0.0, 0.08)\), \(C=(1.07, 0.002, 0.77)\), and \(D=(1.06, 0.01, 0.065)\).  \label{y1y3Rcon}}
\end{figure}
The fit depends on  a total of six coefficients, as can be seen from examining the undiagonalised effective mass matrix. Optimising \(R\), we should also attempt to find mixing angles in line with those known to parameterize the neutrino sector - i.e. large \(\theta_{23}\) and \(\theta_{12}\), with a comparatively small (but non-zero) \(\theta_{13}\). This is necessary to obtain results compatible with neutrino oscillation experiments. Table \ref{nures} summarises the neutrino parameters the model must be in keeping with in order to be acceptable. We should note that the parameter \(m_0\) will be trivially matched up with the mass differences shown in Table \ref{nures}.\\

If we take some choice values of three of our five free parameters, we can construct a contour plot for curves with constant R using the other two. Figure \ref{y1y3Rcon} shows this for a series of fixed parameters. Each of the lines is for \(R=32\), so we can see that there is a deal of flexibility in the parameter space for finding allowed values of the ratio. \\

\begin{figure}[t!]\centering\includegraphics[scale=0.4]{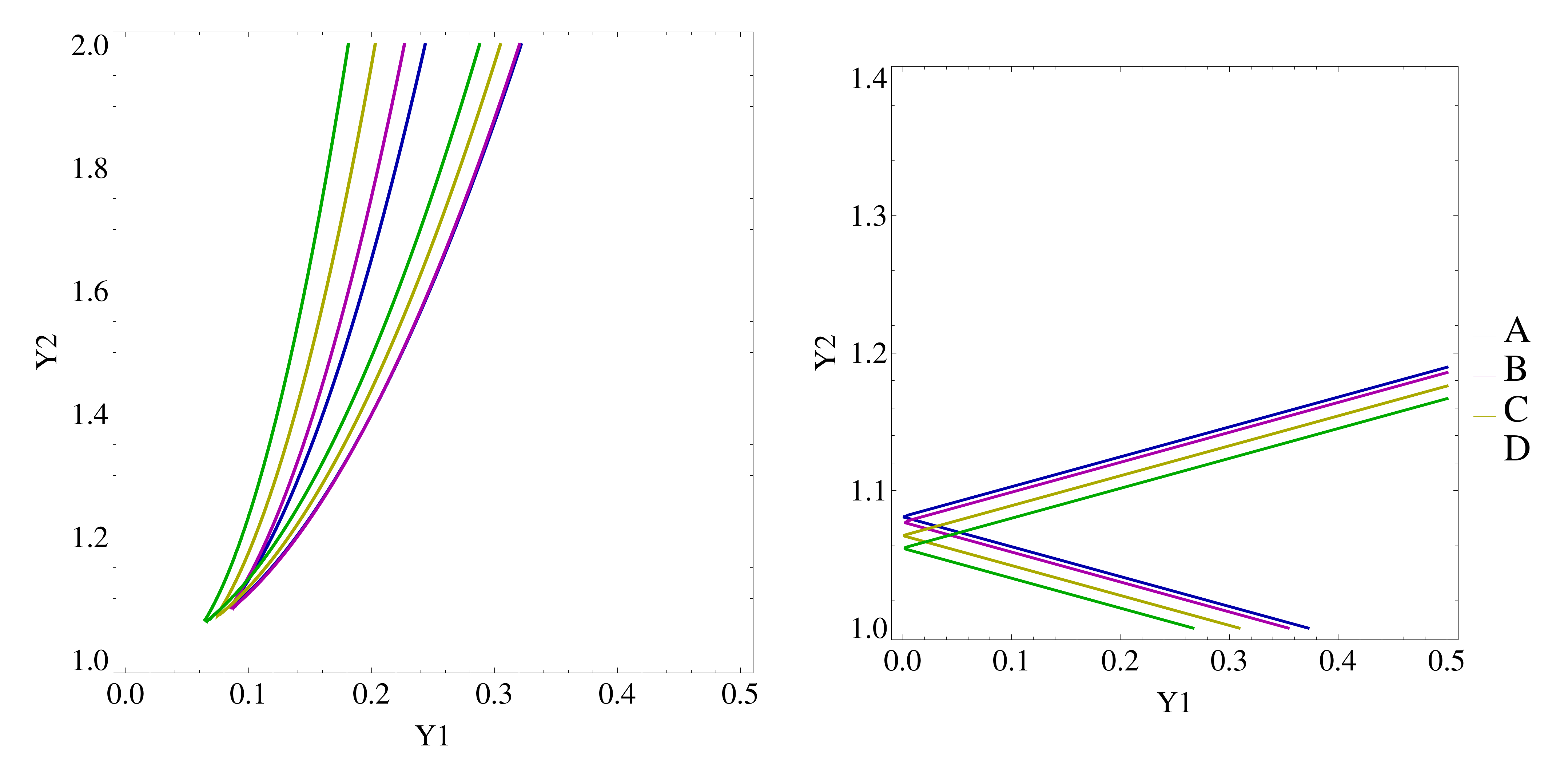}\caption{The figures show plots of two large neutrino mixing angles at their current best fit values. Left: Plot of \(\sin^2(\theta_{12})=0.306\), Right:  Plot of \(\sin^2(\theta_{23})=0.446\). The curves have \((Y_3, Z_1, Z_2)\) values set as follows: \(A=(1.08, 0.05, 0.02)\), \(B=(1.08, 0.0, 0.08)\), \(C=(1.07, 0.002, 0.77)\), and \(D=(1.06, 0.01, 0.065)\). \label{s212}}

\end{figure}•
In order to further determine which parts of the broad parameter space are most suitable for returning phenomenologically acceptable neutrino parameters, we can plot the value of \(\sin^2(\theta_{12})\) or \(\sin^2(\theta_{23})\) in the same parameter space as Figure \ref{y1y3Rcon} - \((Y_1,Y_2)\). The first plot in Figure \ref{s212} shows that the angle \(\theta_{12}\) constraints are best satisfied at lower values of \(Y_1\), while there are the each line spans a large part of the \(Y_2\) space. The second plot of Figure \ref{s212} suggests a preference for comparatively small values of \(Y_2\) based on the constraints on \(\theta_{23}\). As such, we might expect that for this corner of the parameter space there will be some solutions that satisfy all the constraints. \\

\begin{figure}[t!]\centering
\includegraphics[scale=0.45]{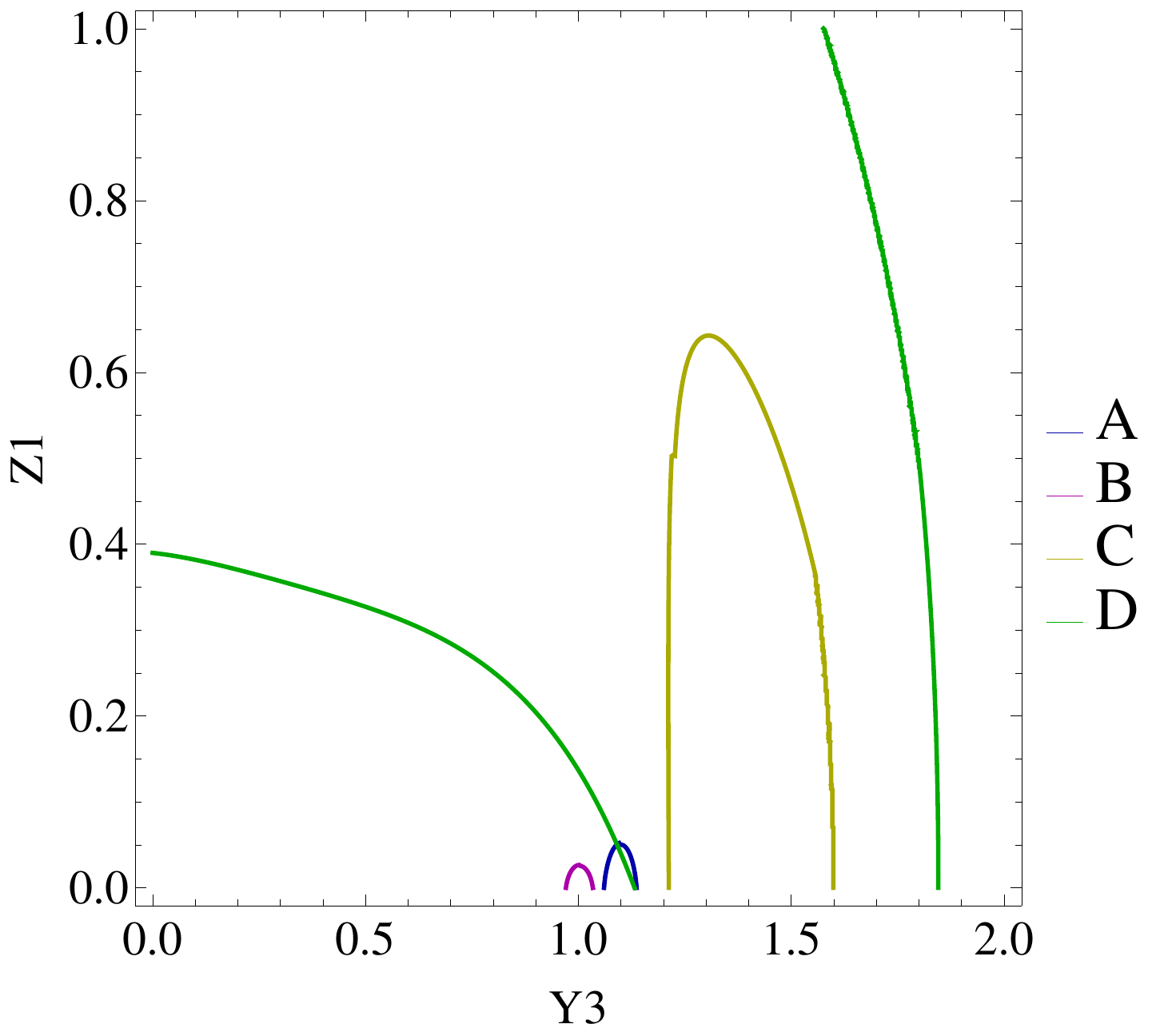}\caption{Plots of lines with the best fit value of \(R=32\) in the parameter space of \((Y_3,Z_1)\). The curves have \((Y_1, Y_2, Z_2)\) values set as follows: \(A=(\frac{1}{5}, 1.4, 0.02)\), \(B=(0.05, 1.5, 0.01)\), \(C=(\frac{1}{2}, 1.6, 0.01)\), and \(D=(\frac{2}{3}, 1.8, 0.5)\).\label{y4z1con}}
\end{figure}

Figure \ref{y4z1con} also shows a plot for contours of best fitting values of  \(R\), with the free variables  chosen as \(Y_3\) and \(Z_1\). As before, this shows that for a range of the other parameters, we can usually find suitable values of \((Y_3,Z_1)\) that satisfy the constraints on \(R\). This being the case, we expect that it should be possible to find benchmark points that will allow for the other constraints to also be satisfied.\\

This flexibility in the parameter space translates to the other experimental parameters, such that the points that allow experimentally allowed solutions are abundant enough that we can fit all the parameters quite well. Table \ref{bm} shows a collection of so-called benchmark points, which are points in the parameter space where all constraints are satisfied within current experimental errors - see Table \ref{nures}. The table only shows values where \(\theta_{23}\) is in the first octant. We might expect that the model should also admit solutions for second octant \(\theta_{23}\), however attempts as numerical solution indicate this possibility is strongly disfavoured. We also note that current Planck data \cite{Ade:2013zuv} puts the sum of neutrino masses to be $\Sigma m_{\nu}\le 0.23\text{eV}$, which the bench mark points are also consistent with. 
\begin{table}[t!]\centering
\begin{tabular}{|c||c|c|c|c|}\hline
Inputs&&&&\\\hline
\(Y_1\)&0.08&0.09&0.09&0.10\\
\(Y_2\)&1.09&1.10&1.10&1.11\\
\(Y_3\)&1.07&1.08&1.08&1.09\\
\(Z_1\)&0.01&0.01&0.00&0.01\\
\(Z_2\)&0.07&0.08&0.08&0.08\\
\(m_0\)&54.0meV&51.6meV&50.3meV&47.8meV\\\hline
Outputs&&&&\\\hline
\(\theta_{12}\)&33.5&33.2&33.1&32.8\\
\(\theta_{13}\)&8.70&8.82&9.05&9.05\\
\(\theta_{23}\)&41.9&41.7&41.7&41.5\\
\(m_1\)&53.4meV&51.1meV&49.8meV&47.3meV\\
\(m_2\)&54.1meV&51.8meV&50.5meV&48.1meV\\
\(m_3\)&73.2meV&71.5meV&70.8meV&69.1meV\\\hline
\end{tabular}
\caption{Table of Benchmark values in the Parameter space, where all experimental constraints are satisfied within errors. These point are samples of the space of all possible points, where we assume \(\theta_{23}\) is in the first octant. All inputs are given to two decimal places, while the outputs are given to 3s.f.  \label{bm}}
\end{table}

\subsection{Proton decay}
Proton decay is a recurring problem in many $SU(5)$ GUT models, with the \textquotedblleft dangerous" dimension six operators, with the effective operator form:
\be\ \frac{QQQL}{\Lambda^2}\,,\,\frac{d^cu^cu^ce^c}{\Lambda^2}\,,\, \frac{\bar{e}^c\bar{u}^cQQ}{\Lambda^2}\,,\,\frac{\bar{d}^c\bar{u}^cQL}{\Lambda^2}\,. \ee
Since there are strong bounds on the proton lifetime ($\tau_p\ge10^33\text{yr}$) then these operators should be highly suppressed or not allowed in any GUT model. \\

Within the context of the $SU(5)\times A_4\times U(1)$ in F-theory, these operators arise from effective operators of the type:
\be 10\cdot10\cdot10\cdot\bar{5}\,, \ee
where the $\bar{5}$ contains the $SU(2)$ Lepton doublet and the $d^c$, and the quark doublet, $u^c$ and $e^c$ arise from $10$ of $SU(5)$. The interaction will be mediated by the $H_u$ and $H_d$ doublets. \\

In the model under consideration, two matter curves are in the $10$ representation of the GUT group: $T_3$ containing the third generation, and $T$ containing the lighter two generations. In general these can be expressed as:
\be\bg T^i\cdot T^j_3\cdot F \\ i+j=3\,\,\text{and}\,\,i,j\in\{0,1,2,3\}\eg \ee
Here, the role of R-symmetry in the model becomes important, since due to the assignment of this symmetry, these operators are all disallowed. Further more, the operators which have $i\neq0$ will have net charge due to the $U(1)_\perp$, requiring them to have flavons to balance the charge. This would offer further suppression in the event that R-symmetry were not enforced. \\

There are also proton decay operators mediated by $D$-Higgs triplets and their anti-particles, which arise from the same operators, but in a similar way, these will be disallowed by R-symmetry thus preventing proton decay via dimension six operators. \\

The dimension four operators, which are mediated by superpartners of the Standard Model, will also be prevented by R-symmetry. However, even in the absence of this symmetry, the need to balance the charge of the $U(1)_{\perp}$ would lead to the presence of additional GUT group singlets in the operators, leading to further, strong suppression of the operator. 

\subsection{Unification}
The spectrum in Table 4 is equivalent to three families of quarks and
leptons plus
three families of $5+\overline{5}$ representations which include the two
Higgs doublets that get VEVs.
Such a spectrum does not by itself lead to gauge coupling unification at
the field theory level, and the splittings which may be present in F-theory
cannot be sufficiently large to allow for unification, as discussed in
\cite{Callaghan:2013kaa}.
However, as discussed in \cite{Callaghan:2013kaa}, where the low energy
spectrum is identical to this model
(although achieved in a different way)  there may be additional bulk
exotics which are capable of restoring gauge coupling unification
and so unification is certainly possible in this mode. We refer the reader to the literature for a full discussion.


\newpage
\pagebreak

\section{$S_3$  models}

Motivated by phenomenological explorations of the neutrino properties under $S_3$, in this section
 we are interested for $SU(5)$ with $S_3$  discrete symmetry and   its 
 subgroup $Z_3$.
More specifically, we analyse monodromies which induce  the breaking of $SU(5)_{\perp}$
to group factors containing the aforementioned non-abelian discrete group.  Indeed, in this section we encountered two 
such symmetry breaking chains, namely cases $ii$) and $iii)$ of (\ref{3S4chains}). 
With respect to the present point of view, novel features are found for case $iii)$. In the subsequent
we present in brief  case $ii)$ and next we analyse  in detail case  $iii)$.

\subsection{The ${\cal C}_3\times {\cal C}_2$ spectral cover split}

As in the $A_4$ case, because these discrete groups originate form the $SU(5)_{\perp}$
we need to work out the conditions  on the associated coefficients $a_i$.
For ${\cal C}_3\times {\cal C}_2$  split the spectral cover equation is
\be  
P_5=\sum_k b_ks^{5-k}=P_{a}P_b=(a_0+a_1s+a_2s^2+a_3s^3)\,(a_4+a_5s+a_6s^2)\label{su5pepr}
\ee
The equations connecting $b_k$'s with $a_i$'s are of the form $b_k\sim \sum_na_n a_{9-n-k}$, the sum referring to
 appropriate values of $n$ which can be read off from (\ref{su5pepr}) or from Table \ref{table1}.  We recall that the $b_k$
 coefficients are characterised by homologies  $[b_k]=\eta -k\,c_1$.
Using this fact as well as the corresponding equations $b_k(a_i)$ given in the last 
column of Table~\ref{table1}, 
we can determine the corresponding homologies of the $a_i$'s in terms of only
one arbitrary parameter which we may take to be the homology $[a_6]=\chi$.  Furthermore
the constraint  $b_1=a_2a_6+a_3a_5=0$ is solved  by introducing a suitable section $\lambda $  such that
$  a_3=-\lambda\,a_6$ and  $a_2=\lambda\, a_5$.

Apart from the constraint $b_1=0$, there  are no other restrictions on the coefficients $a_i$
 in the case of the $S_3$ symmetry.  If, however, we wish to  reduce the $S_3$  symmetry to $ A_3$ 
(which from the point of view of low energy phenomenology is essentially $Z_3$), additional
conditions should be imposed. In this case the model has an $SU(5)\times Z_3\times U(1)$ symmetry.
As in the case of $A_4$ discussed previously,
in order to derive the constraints on $a_k$'s  for the symmetry reduction  $S_3\to Z_3$  
we compute the discriminant, which turns out to be 
   \be  
   \Delta =\left(a_1^2-4 a_0 a_2\right) a_2^2-27 a_0^2 a_3^2+2 a_1 \left(9 a_0
      a_2-2 a_1^2\right) a_3\label{Dcubic}\,,
      \ee
      and demand $\Delta =\delta^2$.   In analogy with the method followed in $A_4$ 
    we re-organise the terms in powers of the  $x\equiv a_1$:
   \be
   \Delta\to f(x) =-4 a_3 x^3+a_2^2 x^2+18 a_0 a_2 a_3 x-a_0 \left(4 a_2^3+27 a_0
      a_3^2\right)\label{cubic}\,.
      \ee
 First,    we observe that in order to write the above expression as a square, the product
     $a_1a_3$ must be  positive definite  ${\rm sign}(a_1a_3)=+$.   Provided
   this condition is fulfilled, then we  require the vanishing of the discriminant 
$\Delta_f$ of the cubic polynomial $f(x)$, namely:
    \[\Delta_f=-64 a_0 a_3 \left(27 a_0 a_3^2-a_2^3\right){}^3=0\,.\]
  This can occur if  the non-trivial relation $a_2^3=27 a_0 a_3^2 $ holds. Substituting back
  to (\ref{Dcubic})  we find that the condition is fulfilled for  $ a_2^2    \propto  a_1 a_3$. The two
  constraints can be combined to give the simpler ones
  \[
   a_0a_3+a_1a_2=0, \; \;     a_2^2+27a_1a_3=0\]    

The details concerning the spectrum,  homologies and flux restrictions of this model
can be found in~\cite{Dudas:2010zb,Antoniadis:2012yk}.
Identifying $t_{1,2,3}=t_a$ and $t_{4,5}=t_b$ ( due to  monodromies) we 
 distribute the matter and Higgs fields over the curves as follows
\[10_M\equiv 10_{t_b},\; \bar 5_{h_d}\equiv \bar 5_{t_a+t_b},\; 5_{h_u}\equiv 5_{-2t_b},\;
 \bar 5_{2t_a}=\bar  5_M,\] 
and the 
  allowed tree-level couplings with non-trivial $SU(5)$ representations are
\be
\begin{split}
{\cal W}&= y_u\,10_M\,10_M\,\bar 5_{h_u}+y_d\,10_M\,\bar 5_M\,\bar 5_{h_d}
\end{split}
\ee

We have already pointed out that the monodromies organise  the $SU(5)_{GUT}$  singlets $\theta_{ij}$
 obtained from the $24\in SU(5)_{\perp}$
 into  two categories. One class carries $U(1)_i$-charges and they  denoted with  $\theta_{ab}$,
 $\theta_{ba}$ while   the second class  $\theta_{aa},\theta_{bb}$  has no $t_i$-`charges'.  
The KK excitations of the latter could  be identified with the right-handed 
 neutrinos. Notice that in the present model the left handed states of the three families reside on 
 the same matter curve. To  generate flavour and in particular neutrino mixing in this model,  
 one may appeal for example to the mechanism discussed in~\cite{Bouchard:2009bu}.
Detailed phenomenological implications for $Z_3$ models have been discussed elsewhere and will not
be presented here.  Within  the present point of view, novel interesting  features are found in
$3+1+1$ splitting which will be discussed in the next sections.

\subsection{$SU(5)$ spectrum for the $(3,1,1)$  factorisation}

In this case the relevant spectral cover polynomial  splits into three factors according to
\[\sum_{k=0}^5 b_ks^{5-k}=\left(a_4 s^3+a_3 s^2+a_2 s+a_1\right) \left(a_5+s a_6\right) \left(a_7+s a_8\right)\]
We can easily extract the equations  determining the coefficients
$b_k(a_i)$, while the corresponding one  for the homologies reads
\be
[b_k]=\eta-k c_1= [a_l]+[a_m]+[a_n],\;k=0,1,\dots,5,\; k+l+m+n=18, \;l,m,n\le 8\label{split311}\ee
As in the previous case, in order to embed the symmetry in $SU(5)_{\perp}$, the condition $b_1=0$ has to be implemented.

The non-trivial representations are found as follows:
The tenplets are determined by
\[b_5=a_1 a_5 a_7=0\]
As before, the equation for fiveplets is given by 
${\cal R}=b_3^2b_4-b_2b_3b_5+b_0b_5^2=0$.
Substitution of  the relevant equations $b_k=b_k(a_i)$ given in Table~\ref{table1} and the condition $b_1=0$ result in the
  factorisation
\be
\label{Z35split}
\begin{split}
 {\cal R}&=\left(a_1 a_6 a_8+a_2 \left(a_6 a_7+a_5 a_8\right)\right)\times
\left(a_1 a_6+a_5 \left(a_2-c a_5 a_7\right)\right)\\
&\times \left(a_7 \left(a_2-c a_5 a_7\right)+a_1 a_8\right)\times
   \left(a_6 a_7+a_5 a_8\right)
\end{split}
\ee
The four factors determine the homologies of the fiveplets dubbed $5_{a},5_{b},5_{c},5_{d}$
 correspondingly. These, together with the tenplets, are given in Table~\ref{S3mcur}.
   \begin{table}[t] \centering%
\begin{tabular}{|c|c|c|c|c|}
\hline
Curve&equation& homology & $U(1)_Y$&$U(1)_X$\\
\hline
$10_{t_{i}}=10_{a}$& $a_1$& $\eta-3c_1-{\chi}-\psi$&$- N_{\chi}-N_{\psi}$ &$M_{10_{a}}$\\ \hline
$10_{t_{4}}=10_{b} $& $a_5$& $-c_1+\chi$&$ N_{\chi}$ &$M_{10_{b}}$\\ \hline
$5_{-t_{i}-t_{j}}=5_{a}$& $a_1 a_6 a_8+a_2 \left(a_6 a_7+a_5 a_8\right)$& $\eta-3c_1$&$ 0$ &$M_{5_{a}}$\\ \hline
$5_{-t_{i}-t_{4}}=5_{b}$&$ a_1 a_6+a_5 \left(a_2-c a_5 a_7\right)$& $\eta -3c_1-{\psi}$&$ -N_{\psi}$ &$M_{5_{b}}$\\ \hline
$5_{-t_{i}-t_{5}}=5_{c}$& $a_7 \left(a_2-c a_5 a_7\right)+a_1 a_8$& $\eta -3c_1-{\chi}$&$- N_{\chi}$ &$M_{5_{c}}$\\ \hline
$5_{-t_{4}-t_{5}}=5_{d}$& $a_6 a_7+a_5 a_8$& $-c_1+{\chi}+\psi$&$ N_{\chi}+N_{\psi}$ &$M_{5_{d}}$\\ \hline
$10_{t_{5}}=10_{c}$& $a_7$& $-c_1+\psi$&$N_{\psi}$&$M_{10_{c}}$\\ \hline
\end{tabular}%
\caption{Matter curves with their defining equations, homologies,  and multiplicities in the case of (3,1,1) factorisation. }
\label{S3mcur}
\end{table}

\subsection{ ${\cal S}_3$ and ${\cal Z}_3 $ models for $(3,1,1)$ factorisation}

In the following we present one characteristic example of F-theory derived effective
 models when we quotient the theory with a ${\cal S}_3$ monodromy.
As already stated, if no other conditions  are imposed on $a_k$
 this model is considered as an $S_3$  variant of the $3+1+1$ example given in~\cite{Dudas:2010zb,Antoniadis:2012yk}.
 In this case the $ 10_{t_i}, i=1,2,3$ residing on a curve - characterised by a common defining equation $a_1=0$ -
 are organised in two irreducible $S_3$ representations $2+1$. 
 The same reasoning  applies  to the remaining representations.
 In Table~\ref{SU5Z3specA_1} we present the spectrum of a model with $N_{\chi}=-1$ and  $N_{\psi}=0$.
Because singlets play a vital role, here, in addition
we include the singlet field spectrum.  Notice that the multiplicities of $\theta_{i4},\theta_{4i}$
 are not determined by  the $U(1)$  fluxes assumed here, hence they are treated
 as free parameters.

   \begin{table}[tb] \centering%
\begin{tabular}{|l|c|c|l|}
\hline
$\quad\quad{SU(5)\times{S_{3}}}$& $N_Y$-flux&$M_X$&Matter\\
\hline
$ 10_{a}^{(1)}=(10,1)_M$&$+1$ &$1$&$Q+2e^c$\\ 
$ 10_{a}^{(2)}=(10,2)_M$&$0$ &$1$&$2(Q+u^{c}+e^c)$\\ \hline
$ \ov{10}_{b}=(\ov{10},1)$& $-1$ &$0$&$u^{c}+\bar e^c$\\ \hline
$ 5^{(1)}_{a}=(5,1)$&$ 0$ &$1$&$h_u+D$\\
$ 5^{(2)}_{a}=(5,2)$&$ 0$ &$1$&$2(h_u'+D')$\\ \hline
$\ov{5}^{(1)}_{b}=(\ov{5},1)_{M}$& $ 0$ &$-1$&$\ell+d^c$\\ 
$\ov{5}^{(2)}_{b}=(\ov{5},2)_{M}$& $ 0$ &$-1$&$2(\ell+d^c)$\\ \hline
$\ov{5}^{(1)}_{c}=(\ov{5},1)$& $1 $ &$-1$&$\bar D$\\ 
$\ov{5}^{(2)}_{c}=(\ov{5},2)$& $ 0$ &$-1$&$2(h_d'+\bar D')$\\ \hline
$ \bar 5_{d}=(\ov{5},1)$& $ -1$ &$0$&$h_d$\\ \hline
$ 10_{c}$&$0$&$0$&empty\\ \hline
\end{tabular}%
\caption{Matter content for an $SU(5)_{GUT}\times{S_{3}}\times{U(1)}$. $S_{3}$ monodromy 
organises  $10_{a}$,$5_{a}$,$5_{b}$ and $5_{c}$ representations in doublets and singlets. }
\label{SU5Z3specA_1}
\end{table}

\pagebreak

\subsection{The Yukawa matrices in $S_3$ Models }

To construct the mass matrices in the case of $S_3$ models we first recall a few useful properties.
There are six elements of 
 the group in three classes, and their irreducible representations are $\textbf{1}$, 
 $\textbf{1}^\prime$ and $\textbf{2}$. The tensor product of 
two doublets, in the real representation, contains two singlets and a doublet:

\begin{equation}
{\bf 2\otimes{2}}={\bf 1}\oplus{\bf{1}^\prime}\oplus{\bf 2} 
\label{S3product}
\end{equation}

\noindent Thus, if $(x_1, x_2)$ and $(y_1, y_2)$ represent the components of the doublets, the above product gives 

\be
{\bf 1}:(x_1 y_1+x_2 y_2),\quad{{\bf{1}^\prime}:(x_1 y_2-x_2 y_1)},
\quad{{\bf 2}:\left(\begin{array}{l}x_1 y_2+x_2 y_1\\x_1 y_1-x_2 y_2\end{array}\right)}.
\label{S3coeff}
\ee

\noindent The singlets are muliplied according to
the rules: ${\bf 1}\otimes{{\bf 1}^\prime}=\bf{1}^\prime$ 
and ${\bf 1}^\prime\otimes{{\bf 1}^\prime}=\bf{1}$. Note that  $\textbf{1}^\prime$ is not an $S_3$ invariant.
With these simple rules in mind, we proceed with the construction of the fermion mass matrices, starting from the quark sector.

\subsection{Quark sector}

We start our analysis of the Top-type quarks. We see from table \ref{SU5Z3specA_1} that we have two types of operators contribute to the Top-type quark matrix.

\noindent 1) A tree level coupling: $g10_{a}^{(2)}\cdot{10_{a}^{(2)}}\cdot{5_{a}^{(1)}}$

\noindent 2) Dimension 4 operators: $\lambda_{1}10_{a}^{(1)}\cdot{10_{b}^{(1)}}\cdot{5_{a}^{(1)}}\cdot{\theta_{a}^{(1)}}$ and 
$\lambda_{2}10_{a}^{(2)}\cdot{10_{b}^{(1)}}\cdot{5_{a}^{(1)}}\cdot{\theta_{a}^{(2)}}$

In order to generate a hierarchical mass spectrum we accommodate the charm and top quarks in the $10_{a}^{(2)}$ curve and the first generation  on the $10_{a}^{(1)}$ curve. In this case, only the first (tree level) coupling contributes to the Top quark terms. Using the $S_{3}$ algebra above while choosing $\langle{5_{a}^{1}}\rangle=\langle{H_{u}}\rangle=\upsilon_{u}$ and $\langle{\theta_{a}^{1}}\rangle=\theta_{0}$, $\langle{\theta_{a}^{2}}\rangle=(\theta_{1},0)^{\text{T}}$ we obtain
 the following mass matrix for the Top-quarks

\be
m_{u}={\left(\begin{array}{ccc}
\lambda_{1}\theta_{0} & \lambda_{2}\theta_{1} & 0 \\ 
0 & \epsilon g & 0 \\ 
0 & 0 & 
g\end{array}\right)\upsilon_{u}}
\label{mup}
\ee

Because two generations live on the same matter curve ($10_{a}^{(2)}$ curve) we implement  the Rank theorem.
For this reason we have suppressed the element-22 in the matrix above with a small scale parameter $\epsilon$. 
The quark eigenmasses are obtained from $V_{u}^{L\dagger}m_{u}m_{u}^{\dagger}V_{u}^{L}=(m_{u}^{diag})^{2}$ where the transformation matrix $V_{u}^{L}$ is required for CKM-matrix along with the transformation of Bottom-type quark masses $V_{d}^{L}$ such that $V_{CKM}=V_{u}^{L\dagger}V_{d}^{L}$. By setting $x=\lambda_{1}\theta_{0}$, $y=\lambda_{2}\theta_{1}$ and $g=z$ we have

\be
m_{u}m_{u}^{\dagger}={\left(\begin{array}{ccc}
x^{2}+y^{2} & \epsilon yz & 0 \\ 
\epsilon yz & \epsilon^{2}z^{2} & 0 \\ 
0 & 0 & 
z^{2}\end{array}\right)\upsilon_{u}^{2}}
\label{mup2}
\ee 

\noindent For reasonable values of the parameters this matrix  leads to mass eigenvalues with 
the required mass hierarchy and a Cabbibo mixing angle. The smaller mixing angles are expected
to be generated from the down quark mass matrix.  Indeed,  the following Yukawa couplings emerge
for the Bottom-type quarks:

\noindent 1) First generation: $g_{1}10_{a}^{(1)}\cdot{\bar{5}_{b}^{(1)}}\cdot{\bar{5}_{d}}\cdot{\theta_{a}^{(1)}}$.

\noindent 2) Second and third generation: $g_{2}10_{a}^{(2)}\cdot{\bar{5}_{b}^{(2)}}\cdot{\bar{5}_{d}}\cdot{\theta_{a}^{(1)}}$.

\noindent 3) First-second, third generation: $g_{3}10_{a}^{(2)}\cdot{\bar{5}_{b}^{(1)}}\cdot{\bar{5}_{d}}\cdot{\theta_{a}^{(2)}}$ and $g_{4}10_{a}^{(1)}\cdot{\bar{5}_{b}^{(2)}}\cdot{\bar{5}_{d}}\cdot{\theta_{a}^{(2)}}$.

\noindent 4) Second-third generation: $g_{5}10_{a}^{(2)}\cdot{\bar{5}_{b}^{(2)}}\cdot{\bar{5}_{d}}\cdot{\theta_{a}^{(2)}}$.

\noindent We assume that the doublet $H_{d}\in {\bar{5}_{b}^{(1)}}$ and the singlet ${\theta_{a}^{2}}$ 
(being a doublet under $S_3$) develop VEVs designated as $\langle{H_{d}\rangle}=\upsilon_{d}$ and  $\langle{\theta_{a}^{2}}\rangle=(\theta_{1},\theta_{2})^{\text{T}}$.  Then, applying the $S_{3}$ algebra,   the Yukawa couplings above  
induce the following mass matrix for the Bottom-type quarks:

\be
m_{d}={\left(\begin{array}{ccc}
g_{1}\theta_{0} & g_{3}\theta_{1} & g_{3}\theta_{2} \\ 
g_{4}\theta_{1} & g_{2}\theta_{0}+g_{5}\theta_{2} & g_{5}\theta_{1} \\ 
g_{4}\theta_{2} & g_{5}\theta_{1} & g_{2}\theta_{0}-g_{5}\theta_{2}\end{array}\right)\upsilon_{d}}.
\label{mdown}
\ee
For appropriate Singlet VEVs the structure  of the Bottom quark mass matrix is capable to 
reproduce the hierarchical mass spectrum and the required CKM mixing.

\subsection{Leptons}

The charged leptons will have the same couplings as the Bottom-type quarks. To simplify the analysis,
let us start with a simple case where the Singlet VEVs exhibit the hierarchy $\theta_2<\theta_1<\theta_0$.
Furthermore, taking  the limit   $\theta_{2}\rightarrow{0}$ and switching-off the Yukawas coefficients 
$g_{3}$, $g_{4}$ in (\ref{mdown}) we achieve a block diagonal form of the charged lepton matrix

 \be
m_{\ell}={\left(\begin{array}{ccc}
g_{1}\theta_{0} & 0 & 0 \\ 
0 & g_{2}\theta_{0} & g_{5}\theta_{1} \\ 
0 & g_{5}\theta_{1} & g_{2}\theta_{0}\end{array}\right)\upsilon_{d}}.
\label{mleptonblock}
\ee
with eigenvalues
\be
\begin{split}
m_{e}=g_{1}\theta_{0},\;
m_{\mu}=g_{2}\theta_{0}-g_{5}\theta_{1},\;
m_{\tau}=g_{2}\theta_{0}+g_{5}\theta_{1}
\end{split}
\label{mutaublock}
\ee
and maximal mixing between the second and third generations.

We turn now our attention to the couplings of the neutrinos. We identify the right-handed neutrinos with the SU(5)-singlet $\theta_{c}=1_{ij}$. Under the $S_{3}$ symmetry, $\theta_{c}$ splits into a singlet, named $\theta_{c}^{(1)}$ and a doublet, $\theta_{c}^{(2)}$. As in the case of the quarks and the charged leptons we distribute the right handed neutrino species as follows

\[\theta_{c}^{(1)}\rightarrow{\nu_{1}^{c}}\quad\quad{\textrm{and}}\quad\quad\theta_{c}^{(2)}
\rightarrow{(\nu_{2}^{c}\quad{\nu_{3}^{c}})^{\text{T}}}\]

The Dirac neutrino mass matrix arises from the following couplings 

\noindent 1) $y_{1}5_{a}^{(1)}\cdot{\bar{5}_{b}^{(1)}}\cdot{\theta_{c}^{(1)}}\cdot{\theta_{a}^{(1)}}$

\noindent 2) $y_{2}5_{a}^{(1)}\cdot{\bar{5}_{b}^{(2)}}\cdot{\theta_{c}^{(2)}}\cdot{\theta_{a}^{(1)}}$

\noindent 3) $y_{3}5_{a}^{(1)}\cdot{\bar{5}_{b}^{(2)}}\cdot{\theta_{c}^{(1)}}\cdot{\theta_{a}^{(2)}}$

\noindent 4) $y_{4}5_{a}^{(1)}\cdot{\bar{5}_{b}^{(1)}}\cdot{\theta_{c}^{(2)}}\cdot{\theta_{a}^{(2)}}$

\noindent 5) $y_{5}5_{a}^{(1)}\cdot{\bar{5}_{b}^{(2)}}\cdot{\theta_{c}^{(2)}}\cdot{\theta_{a}^{(2)}}$

\noindent and has the following form (for $\theta_{2}\rightarrow{0}$)

\be
{\cal M}_{D}={\left(\begin{array}{ccc}
y_{1}\theta_{0} & y_{3}\theta_{1} & 0 \\ 
y_{4}\theta_{1} & y_{2}\theta_{0} & y_{5}\theta_{1} \\ 
0 & y_{5}\theta_{1} & 
y_{2}\theta_{0}\end{array}\right)\upsilon_{u}}
\label{mDir}
\ee

Although  the Dirac mass matrix has the same form with the charged lepton matrix (\ref{mdown}) 
in general they have  different Yukawas coefficients. Thus, substantial mixing effects may also
occur even in the case of a diagonal heavy Majorana mass matrix.

In the following we construct  effective neutrino mass matrices  compatible with the well known neutrino
data in two different ways.
In the first approach we take the simplest scenario for a diagonal heavy Majorana  mass matrix and  generate the
 TB-mixing combining charged lepton and neutrino block-diagonal textures. In the second case we consider the
  most general form of the Majorana matrix and we try to generate TB-mixing only from the Neutrino sector.

\subsubsection{Block diagonal case}

We start with the attempt to generate the TB-mixing combining charged lepton and neutrino block-diagonal
textures. The Majorana matrix will simply be the identity matrix scaled by a RH-neutrino mass $M$.
The effective neutrino mass matrix \(M_{eff}=M_DM_{M}^{-1}M_D^{\text{T}}\) now reads:

\be
{\cal M}_{\nu}^{eff}=\left(\begin{array}{ccc}
y_{1}^{2}\theta_{0}^{2}+y_{3}^{2}\theta_{1}^{2} & (y_{2}y_{3}+y_{1}y_{4})\theta_{0}\theta_{1} &  y_{3}y_{5}\theta_{1}^{2} \\ 
 (y_{2}y_{3}+y_{1}y_{4})\theta_{0}\theta_{1} & y_{2}^{2}\theta_{0}^{2}+(y_{4}^{2}+y_{5}^{2})\theta_{1}^{2}  & 2y_{2}y_{5}\theta_{0}\theta_{1} \\ 
 y_{3}y_{5}\theta_{1}^{2}  & 2y_{2}y_{5}\theta_{0}\theta_{1} & y_{2}^{2}\theta_{0}^{2}\end{array}\right)\frac{\upsilon_{u}^{2}}{M}
\ee

\noindent where we used the Dirac mass matrix as given in (\ref{mDir}). First of all we observe that we can reduce the number of the parameters by defining \[   \theta_0 =\epsilon \theta_1,\; x=y_2\epsilon,\; y=y_1\epsilon,\; a=y_3,\; b=y_4,\; c=y_{5}.\]
Then ${\cal M}_{\nu}^{eff}$ is written
\be
{\cal M}_{\nu}^{eff}=\left(\begin{array}{ccc}
y^{2}+a^{2} & xa+yb &  ac \\ 
xa+yb  & x^2+b^{2}+c^2 & 2xc \\ 
ac & 2xc & x^2\end{array}\right)\frac{\upsilon_{u}^{2}\theta_{1}^{2}}{M}
\label{mYukawass}
\ee 

In the limit of a small  $y_{5}$ Yukawa  (or  $c\rightarrow{0}$)  we achieve a block diagonal form  given by

\be
{\cal M}_{\nu}^{eff}=\left(\begin{array}{ccc}
y^{2}+a^{2} & xa+yb &  0 \\ 
xa+yb  & x^2+b^{2} & 0 \\ 
0 & 0 & x^2\end{array}\right)\frac{\upsilon_{u}^{2}\theta_{1}^{2}}{M}
\label{m0Yukawass}
\ee 
This can be diagonalised by a unitary matrix
\be
V_{\nu}=\left(
\begin{array}{ccc}
 \cos (\theta_{12}) & \sin (\theta_{12}) & 0 \\
 -\sin (\theta_{12}) & \cos (\theta_{12}) & 0 \\
 0 & 0 & 1\\
\end{array}
\right)
\ee  
Now, we may  appeal to the  block diagonal form of the charged lepton matrix  
(\ref{mleptonblock}) which introduces a maximal  $\theta_{23}$ angle  so that the final mixing is
\[U_{eff}=\left(
\begin{array}{ccc}
 \cos (\theta_{12}) & \sin (\theta_{12}) & 0 \\
 -\cos (\theta_{23}) \sin (\theta_{12}) & \cos (\theta_{12}) \cos (\theta_{23}) & \sin (\theta_{23}) \\
 \sin (\theta_{12}) \sin (\theta_{23}) & -\cos (\theta_{12}) \sin (\theta_{23}) & \cos (\theta_{23}) \\
\end{array}
\right)
\]

\noindent Indeed, a quick calculation in the 2-3 block of charged lepton matrix (\ref{mleptonblock}) gives:

\[ \cos{(2\theta_{23})}=0\rightarrow{\theta_{23}=\frac{\pi}{4}}\]

Moreover, diagonalisation of  the neutrino mass matrix  yields

\be
\tan(2\theta_{12})=\frac{2(x\alpha+y b)}{y^{2}+\alpha^{2}-x^{2}-b^{2}}
\ee

\begin{figure}[!t]
  \centering
  \includegraphics[scale=0.4]{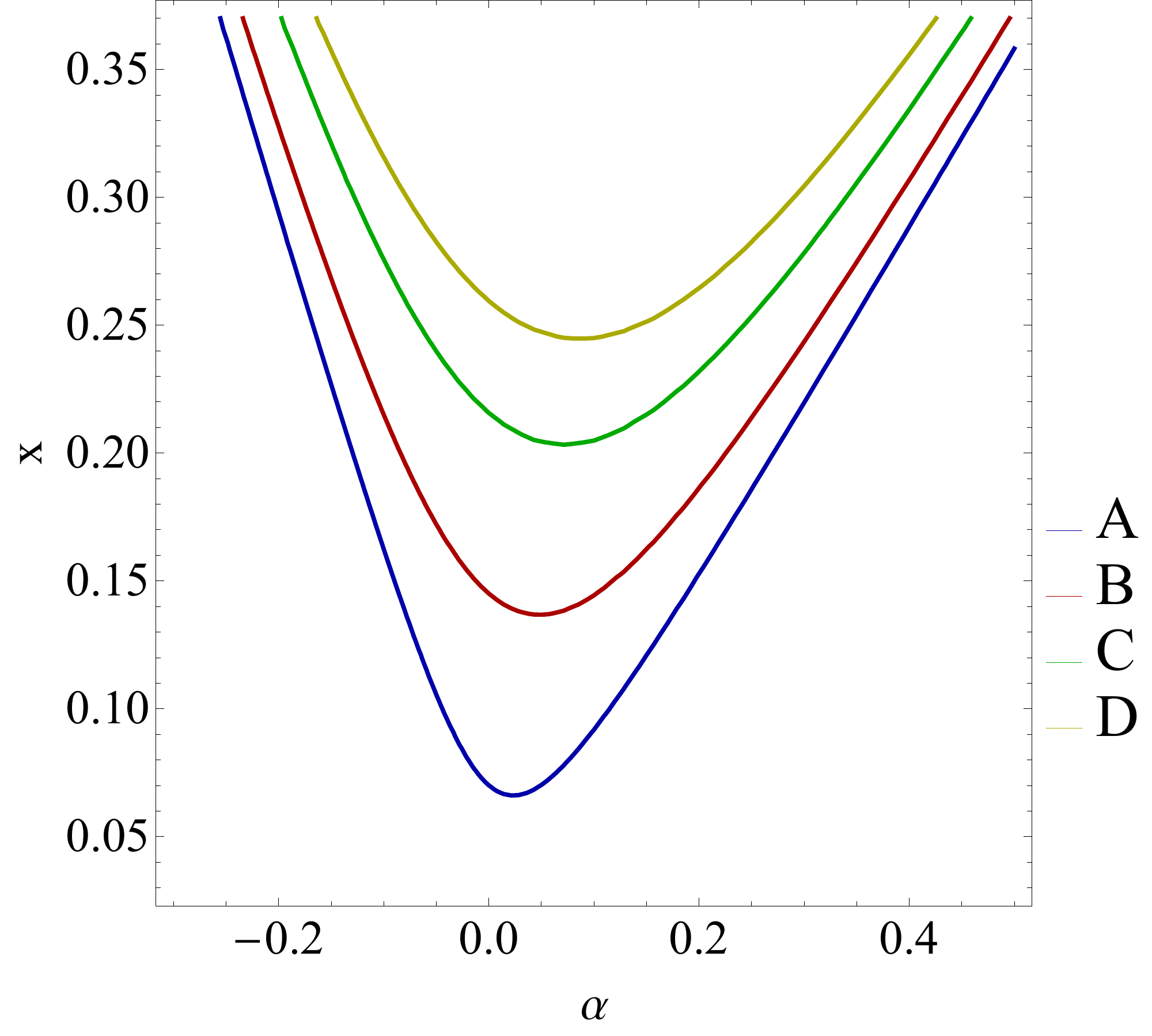}
  \caption{Curves for the relation $\tan{(2\theta_{12})}=2.828$ in the parameter space of $(\alpha,x)$ . The curves 
  have (b,y) values set as follows: $A=(\frac{2}{5},\frac{4}{7})$, $B=(\frac{1}{3},\frac{1}{2})$, $C=(\frac{4}{13},\frac{1}{2})$ and $D=(\frac{2}{7},\frac{1}{2})$.  }
  \label{tan2phi}
  \end{figure}

The TB-mixing matrix now arises for $\tan{(2\theta_{12})}\approx{2.828}$. In figure (\ref{tan2phi}) 
we plot contours for the above relation in the plane ($\alpha, x$) for various values of the pairs $(b,y)$.As can be observed,  $\tan{(2\theta_{12})}$ takes the desired value for reasonable range of the parameters
 $\alpha,b,x,y$. For example

\be
\tan{(2\theta_{12})}\approx{2.804}\quad\quad{for}\quad\quad(\alpha,b,x,y)=(\frac{2}{7},\frac{2}{9},\frac{1}{4},\frac{3}{8})
\label{parameters}
\ee
We conclude that  the simplified  (block-diagonal) forms of the charged lepton and neutrino mass matrices are 
compatible with the TB-mixing. 
It is easy now to obtain the known deviations of the TB-mixing allowing small values for the parameters $c, \theta_2$
in  (\ref{mleptonblock}) and  (\ref{m0Yukawass}) respectively. However, we also need to reconcile the
 ratio of the mass square differences $R=\Delta m^{2}_{32}/\Delta m^{2}_{21}$ with the experimental data $R\approx 32$. 
 To this end, we first compute the mass eigenvalues of the effective neutrino mass matrix

\begin{align*}
m_{1}&=x^{2}\\
m_{2}&=\frac{1}{2}(a^{2}+b^{2}+x^{2}+y^{2}-\Delta)\\
m_{3}&=\frac{1}{2}(a^{2}+b^{2}+x^{2}+y^{2}+\Delta)
\end{align*}

\noindent where $\Delta=\sqrt{[(\alpha+b)^{2}+(x-y)^{2}][(\alpha-b)^{2}+(x+y)^{2}]}$. Notice that $\Delta$ is a positive quantity and as a result $m_{3}>m_{2}$.

\begin{figure}[!t]
  \centering
 \includegraphics[scale=0.6]{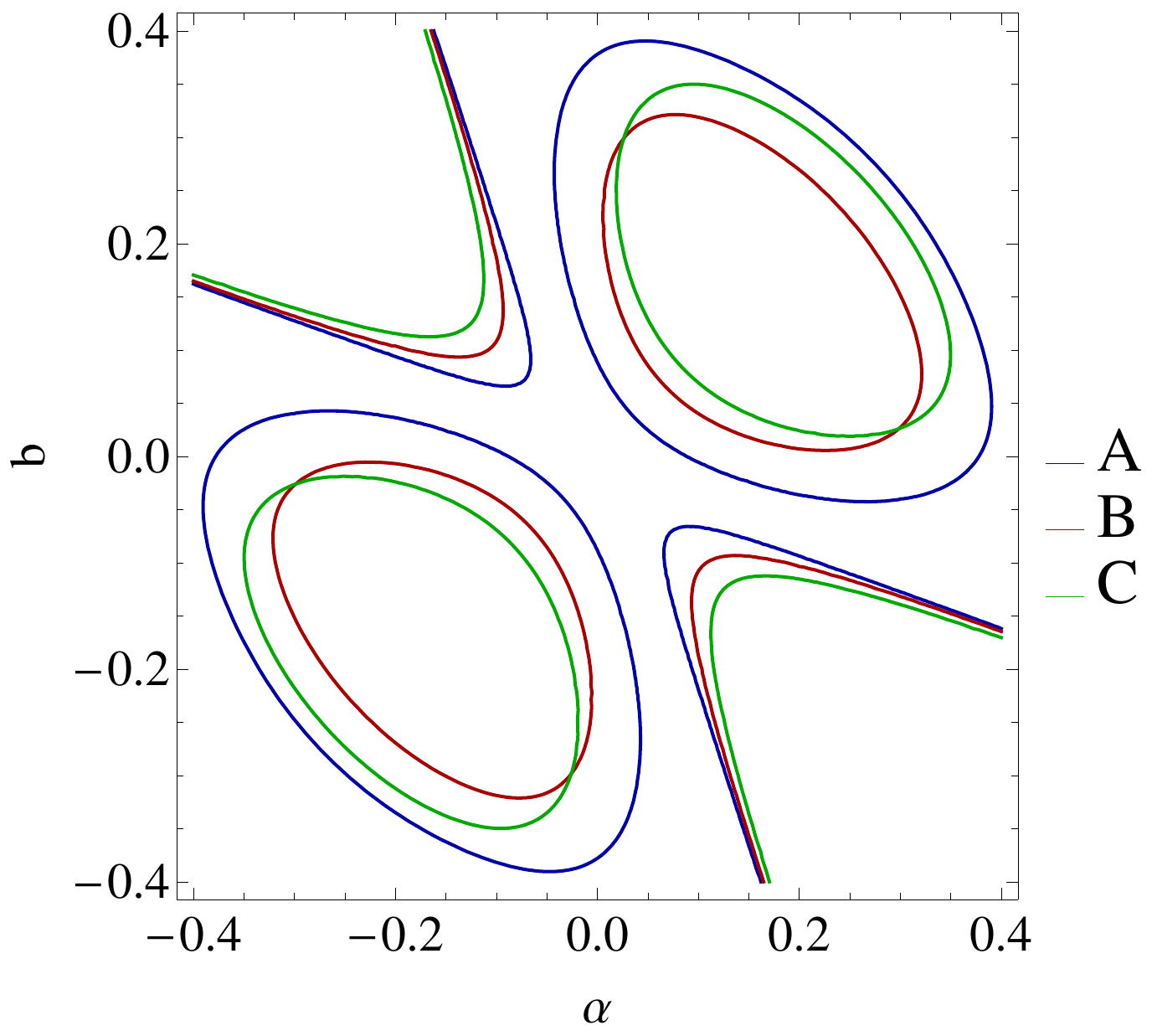}
  \caption{Contour plot for the ratio $R=\Delta m^{2}_{32}/\Delta m^{2}_{21}=32$ in the parameter space $(\alpha, b)$. The curves correspond to the following $(x,y)$ values: $A=(\frac{1}{4},\frac{3}{8})$, $B=(\frac{2}{7},\frac{1}{2})$, and $C=(\frac{2}{7},\frac{3}{7})$ .  }
  \label{dm2ratio}
  \end{figure}

 We can find easily solutions for a wide range of the parameters consistent with the experimental data.  Note that for the same values as in (\ref{parameters}) we achieve a reasonable value of $R\approx{28.16}$.
 In figure(\ref{dm2ratio}) we plot contours of the ratio in the plane ($\alpha$, $b$) for various values of the pair ($x,y$).

We have stressed above that we could generate the $\theta_{13}$ angle by assuming  small values of the Yukawas  $y_5$.
However, this case turns out to be too restrictive since the structure  of (\ref{m0Yukawass}) results to maximal  $(1-2)$ mixing in
contradiction with the experiment. The issue could be remedied by a fine-tuning of the charged lepton mixing, however we 
would like to look up for a natural solution. Therefore, we proceed with other options.

\subsubsection{TB mixing from neutrino sector.}

In the previous analysis we considered the simplest scenario for the Majorana matrix. The general form of the Majorana mass matrix arises by taking into account all the possible flavon terms contributions and has the following form

 \be
 M_{maj}=
\left(
\begin{array}{lll}
 M & f_{1} & f_{2} \\
 f_{1} & m & f_{3} \\
 f_{2} & f_{3} & m
\end{array}
\right) \label{majorana}
\ee

\noindent with $M>m>f_{i}$.

To reduce the number of parameters we consider that $f_{i}=f$ for $i=1,2,3$ and $y_{3}, y_{4}\rightarrow{0}$ in the Dirac matrix.
 In this case the elements of the effective neutrino mass matrix are

\be
\begin{split}
M_{11}&=(a^{2}-b^{2})y^{2}\\
M_{12}&=M_{13}=M_{21}=M_{31}=bx(b-a)(y+c)\\
M_{22}&=M_{33}=(a-b^{2})x^{2}+(a+b^{2})c^{2}-2(b-1)bxc\\
M_{23}&=M_{32}=b(b-1)(x^{2}+c^{2})-2(b^{2}-a)xc\\
\end{split}
\ee

\noindent with an overall factor $\sim\frac{\upsilon_{u}^{2}\theta_{1}^{2}M^{2}}{(2f^{3}-2f^{2}m-f^{2}M+m^{2}M)}$ and the parameters are defined as $a=m/M$, $b=f/M$, $c=y_{5}$, $x=\epsilon y_{2}$, $y=\epsilon y_{1}$ and $\theta_{0}=\epsilon \theta_{1}$. The  matrix 
assumes the general structure:

 \be
 M_{\nu}=
\left(
\begin{array}{lll}
 p & q & q \\
 q & r & s \\
 q & s & r
\end{array}
\right) \label{gen}
\ee

\noindent Maximal atmospheric neutrino mixing and $\theta_{13}=0$ immediately follow from this structure. 
The solar mixing angle  $\theta_{12}$ is not predicted, but it is expected to be large.

Next we try  to generate  TB -mixing only from the neutrino sector (assuming that
the  charged lepton mixing is negligible so that it can be used to lift $\theta_{13}\ne 0$). Then, it is enough to  compare the entries
of the effective mass matrix with the most general mass matrix form which complies with TB-mixing
 \be
 m_{\nu}=
\left(
\begin{array}{lll}
 u & v & v \\
 v & u+w & v-w \\
 v & v-w & u+w
\end{array}
\right) \label{mnTB}
\ee
A quick comparison results to the following simple relations 
\be
\begin{split}
u&=M_{11}\\
v&=M_{12}\\
w&=M_{22}-u=M_{22}-M_{11}\\
\end{split}
\label{uvw}
\ee

\noindent while the (23) element is subject to the constraint:

\be
v=M_{23}+w
\ee

\noindent which results to a quadratic equation of $b$ with solutions being functions of
the remaining parameters  $b=B_{\pm}(a,c,x,y)$.  We choose one of the roots,  $b=B_{-}$,
and substitute it back to the equations (\ref{uvw}) to express the parameters $u$, $v$
 and $w$ as functions of  $(a,c,x,y)$. 

The requirement that all the large mixing effects emerge from the neutrino sector imposes severe restrictions on the parameter space. Hence we need to check their compatibility with the mass square differences ratio $R$. We can 
express the latter as a function of the parameters $R=R(a,c,x,y)$ by noting that the mass eigenvalues 
are given by

\[ m_{1}=u-v,\quad{m_{2}=u+2v},\quad{m_{3}=u-v+2w}\]

\begin{figure}[!t]
  \centering
  \includegraphics[scale=1.2]{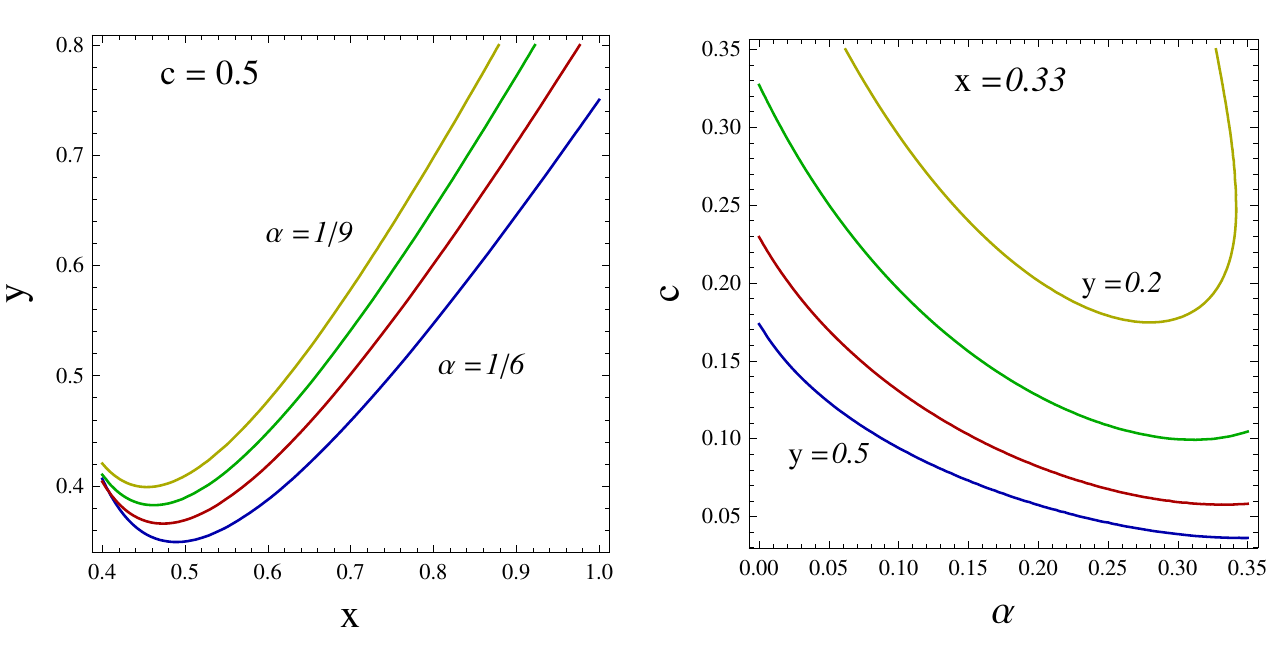}
  \caption{Contour plots for the ratio $R=\Delta m^{2}_{32}/\Delta m^{2}_{21}=32$ in the parameter spaces $(x, y)$-left and $(a,c)$-right. In the first plot (left) $c=0.5$ and $a_{blue}=\frac{1}{6}$, $a_{red}=\frac{1}{7}$, $a_{green}=\frac{1}{8}$ and $a_{yellow}=\frac{1}{9}$. In the $(a,c)$ plot $x=0.33$ and $a_{blue}=0.5$, $a_{red}=0.4$, $a_{green}=0.3$ and $a_{yellow}=0.2$. }
  \label{xyac}
  \end{figure}

Direct substitution gives the desired expression  $R(a,c,x,y)$ which is plotted in figure \ref{xyac}.  It is straightforward to notice that there is a wide range of parameters consistent with the experimental data. In the first graph of the figure we plot contours for the ratio in the plane $(x,y)$ for various values of $a$ and constant value $c=0.5$. In the second graph we plot the ratio in the $(a,c)$ plane with constant $x=0.33$. Note that in both cases, the $a,c,x,y$ parameters take values $<1$.

\begin{figure}[!t]
  \centering
  \includegraphics[scale=0.7]{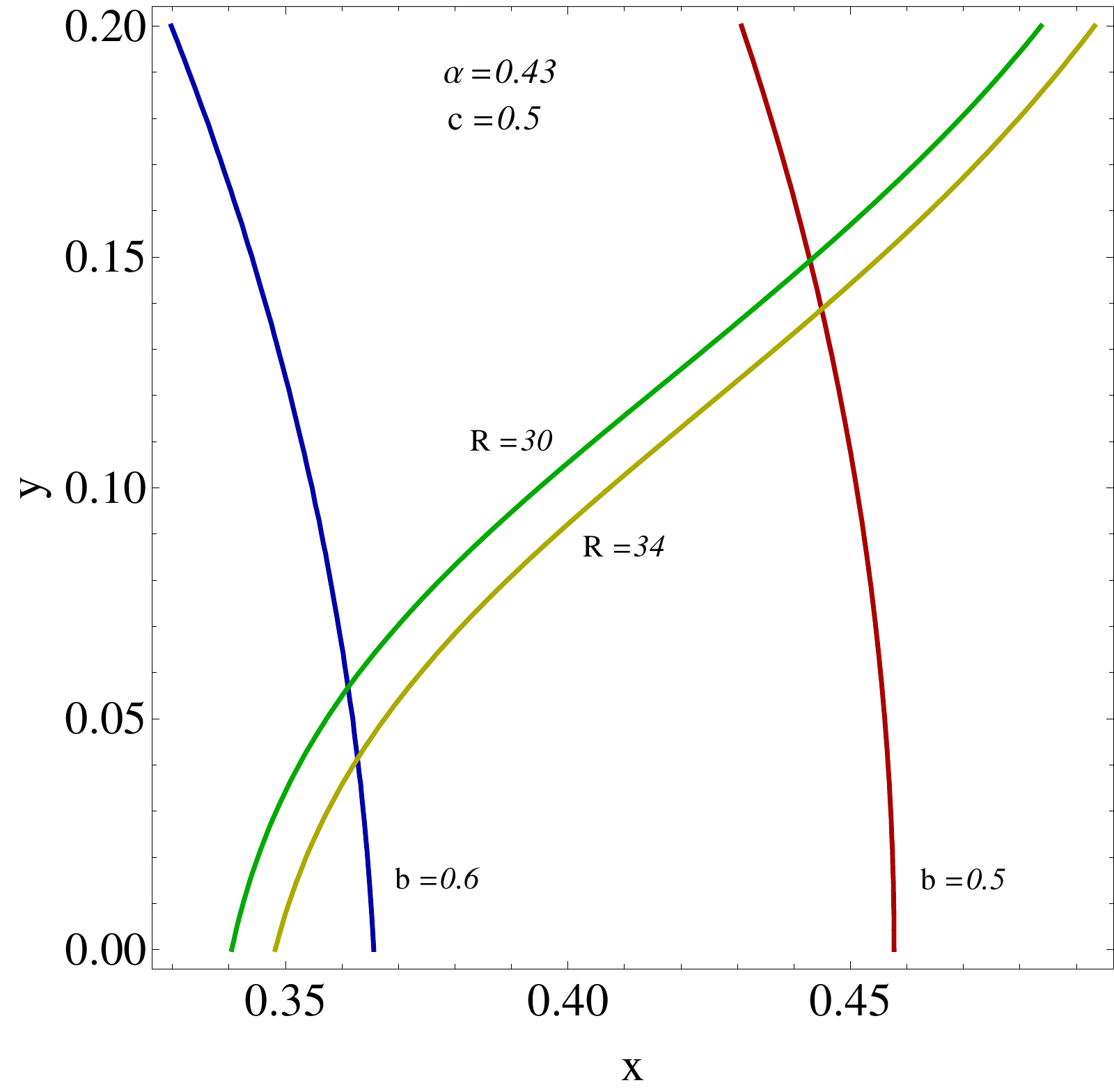}
  \caption{Bounds in the parameter space $(x,y)$ from the experimental data and the requirement $b<1$. }
  \label{bR}
  \end{figure}

Having checked that the parameters $a,c,x,y$ are in the perturbative range, while consistent with the TB-mixing and the mass data, we also should require  that  $b=f/M$ remains in the perturbative regime, i.e. $b<1$.  In figure \ref{bR} we plot the bounds put by this constraint. In particular we plot the mass square ratio in the $(x,y)$ plane for $R=30$ and $R=34$ and we notice that there exists an overlapping region for values of $b$ between 0.5 and 0.6. In this region $x\sim{0.4}$ and $y\sim{0.1}$. More precisely a typical set
of such values gives

\[ (a,c,x,y)=(\frac{3}{7},\frac{1}{2},\frac{2}{5},\frac{1}{10})\rightarrow{b\approx{0.5}}\quad\textrm{and}\quad{R\approx{31.5}}.\]

\section{Conclusions}

\hspace{0.5cm} In this work we considered the phenomenological implications of F-theory $SU(5)$ models with non-abelian discrete family symmetries.
 We discussed the physics of these  constructions
  in the context of the
 spectral cover, which, in the elliptical fibration and under the specific choice of $SU(5)$  GUT,  implies that the discrete family
symmetry must be a subgroup of the permutation symmetry $S_5$.
 Furthermore, we exploited the topological properties of the associated
 5-degree polynomial coefficients (inherited from the internal
 manifold) to derive constraints on the effective field theory models. Since we dealt with discrete gauge groups, we also proposed a discrete
version of the flux mechanism for the  splitting of representations. We started our analysis  splitting appropriately with the spectral cover in order to implement  the $A_4$  discrete symmetry as a subgroup of $S_4$.  Hence, using Galois Theory techniques, we studied the necessary conditions on the discriminant in order to reduce the symmetry from $S_4$ to $A_4$.  Moreover,  we derived the properties of the matter curves accommodating the massless spectrum and the constraints on the Yukawa sector of the effective models.  Then, we first made a  choice  of our flux parameters and picked up a suitable combination  of trivial and non-trivial $A_4$  representations to accommodate the three generations so that a hierarchical mass spectrum for the charged fermion sector is guaranteed.  Next, we focused on the implications of the neutrino sector.  Because of the rich structure of the effective theory emerging from the covering $E_8$ group, we found a considerable number of Yukawa operators contributing to the neutrino mass matrices.  Despite their complexity, it is remarkable that the F-theory constraints and the induced discrete symmetry organise them in a systematic manner so that they accommodate naturally the observed large mixing effects and the smaller $\theta_{13}$ angle of the neutrino mixing matrix.

In the second part of the present article, using the appropriate factorisation of the spectral cover we derive the $S_3$ group as a family symmetry which accompanies the $SU(5)$ GUT.  Because now the family symmetry is smaller than before, the resulting fermion mass structures turn out to be less constrained.  In this respect, the $A_4$ symmetry appears to be more predictive.  Nevertheless,  to start with, we choose to focus  on a particular region of the parameter space assuming some of the Yukawa matrix elements are zero and imposing a diagonal heavy Majorana mass matrix. In such cases, we can easily derive block diagonal lepton mass matrices which incorporate large neutrino mixing effects as required by the experimental data.  Next, in a more involved example, we allow for a general Majorana mass matrix and initially determine stable regions of the parameter space which are consistent with TB-mixing.  The tiny $\theta_{13}$ angle can easily arise from small deviations of these values or by charged lepton mixing effects.  Both models derived here satisfy the neutrino mass squared difference ratio predicted by neutrino oscillation experiments.

In conclusion, F-theory  $SU(5)$  models with non-abelian discrete family
symmetries
provide a promising theoretical framework within which the flavour
problem may be addressed.
The present paper presents the first such realistic examples based on
$A_4$ and $S_3$, which are amongst the
most popular discrete symmetries used in the field theory literature in
order to account for neutrino masses and mixing angles.
By formulating such models in the framework of F-theory $SU(5)$, a deeper
understanding of the origin of these discrete symmetries
is obtained, and theoretical issues such as doublet-triplet splitting may
be elegantly addressed.\\\\\\\\

\noindent\textbf{Acknowledgements}\quad

  The research  of AK and GKL has been
co-financed by the European Union (European Social Fund - ESF) and
Greek national funds through the Operational Program "Education and
Lifelong Learning" of the National Strategic Reference Framework
(NSRF) - Research Funding Program: "THALIS". Investing in the
society of knowledge through the European Social Fund. SFK acknowledges the EU ITN grant INVISIBLES 289442. AKM is supported by an STFC studentship.


\newpage
\appendix

\section{Block Diagonalisation of \(A_4\)}\label{a4basis}
\subsection{Four dimensional case}
From considering the symmetry properties of a regular tetrahedron, we can see quite easily that it can be parameterised by four coordinates and its transformations can be decomposed into a mere two generators. If we write these coordinates as a basis for \(A_4\), which is the symmetry group of the tetrahedron, it would be of the form \(\left(t_1,t_2,t_3,t_4\right)^{\text{T}}\). The two generators can then be written in matrix form explicitly as:
\begin{equation}S=\left(\begin{array}{cccc}0&0&0&1\\0&0&1&0\\0&1&0&0\\1&0&0&0\end{array}\right)\,\,\,\text{and}\,\,\,
T=\left(\begin{array}{cccc}1&0&0&0\\0&0&0&1\\0&1&0&0\\0&0&1&0\end{array}\right)\,\,.\end{equation}
However, it is well known that \(A_4\) has an irreducible representation in the form of a singlet and triplet under these generators. If we consider the tetrahedron again, this can be physically interpreted by observing that under any rotation through one of the vertices of the tetrahedron the vertex chosen remains unmoved under the transformation.\footnote{This is a trivial notion for the T generator, but slightly more difficult for the S generator. In the latter case, consider fixing one vertex in place and performing the transformation about it.} In order to find the irreducible representation, we must note some conditions that this decomposition will satisfy. \\

In order to obtain the correct basis, we must find a unitary transformation \(V\) that block diagonalises the generators of the group. As such, we have the following conditions:
\begin{equation}\begin{gathered} VSV^{\text{T}}=S'=\left(\begin{array}{cccc}1&0&0&0\\0&-&-&-\\0&-&-&-\\0&-&-&-\end{array}\right)\,,\\ VTV^{\text{T}}=T'=\left(\begin{array}{cccc}1&0&0&0\\0&-&-&-\\0&-&-&-\\0&-&-&-\end{array}\right)\,,\\VV^{\text{T}}=\text{I}_{4x4}\,\,,\end{gathered}\end{equation}
as well as the usual conditions that must be satisfied by the generators: \(S^2=T^3=(ST)^3=\text{I}\). It will also be useful to observe three extra conditions, which will expedite finding the solution. Namely that the block diagonal of one of the two generators must have zeros on the diagonal to insure the triplet changes within itself. \\

If we write an explicit form for V, 
\begin{equation}V=\left(\begin{array}{cccc}v_{11}&v_{12}&v_{13}&v_{14}\\v_{21}&v_{22}&v_{23}&v_{24}\\v_{31}&v_{32}&v_{33}&v_{34}\\v_{41}&v_{42}&v_{43}&v_{44}\end{array}\right)\,,\end{equation}
we can extract a set of quadratic equations and attempt to solve for the elements of the matrix. Note that we have assumed as a starting point that \(v_{ij}\in\mathbb{R}\forall i,j\). The complete list is included in the appendix. The problem is quite simple, but at the same time would be awkward to solve numerically, so we shall attempt to simplify the problem analytically first. If we start be using: 
\begin{equation}\begin{gathered}v_{11}^2+v_{12}^2+v_{13}^2+v_{14}^2=1\,,\\
\text{\&}\,\,\, 2v_{12}v_{13}+2v_{11}v_{14}=1\,,\end{gathered}\end{equation}
we can trivially see two quadratics,
\begin{equation}(v_{11}-v_{14})^2+(v_{12}-v_{13})^2=0\,.\end{equation}
Since we assume that all our elements or V are real numbers, it must be true then that:
\begin{equation} v_{11}=v_{14} \,\,\,\text{and}\,\,\,v_{12}=v_{13}\,.\end{equation}
We may now substitute this result into a number of equations. However, we chose to focus on the following two:
\begin{equation}\begin{gathered}v_{11}v_{21}+v_{12}v_{23}+v_{13}v_{24}+v_{14}v_{22}\rightarrow v_{11}(v_{21}+v_{22})+v_{12}(v_{23}+v_{24})\,\,\,\text{and}\\
v_{11}v_{21}+v_{12}v_{24}+v_{13}v_{22}+v_{14}v_{23}\rightarrow v_{11}(v_{21}+v_{23})+v_{12}(v_{22}+v_{24})\,.
\end{gathered}\end{equation}
Taking the difference of these two equations, we can easily see there is a solution where \(v_{11}=v_{12}\), and as such by the previous result:
 \begin{equation}v_{11}=v_{12}=v_{13}=v_{14}=\pm\frac{1}{2}\,.\end{equation}
We are free to choose whichever sign for these four elements we please, provided they all have the same sign. This outcome reduces the number of useful equations to twelve, as nine of them can be summarised as 
\begin{equation}\sum_iv_{2i}=\sum_iv_{3i}=\sum_iv_{4i}=0\,.\end{equation}
Let us consider the first of these three derived conditions, along with the conditions:
\begin{equation}\begin{gathered}v_{21}^2+v_{22}^2+v_{23}^2+v_{24}^2=1\,,\\ v_{21}^2+v_{22}v_{23}+v_{22}v_{24}+v_{23}v_{24}=0\,.\end{gathered}\end{equation}
Squaring the condition \(\sum_iv_{2i}=0\) and using these relations, we can derive easily that \(v_{21}=\pm\frac{1}{2}\). Likewise we can derive the same for \(v_{31}\) and \(v_{41}\). As before, we might chose either sign for each of these elements, with each possibility yielding a different outcome for the basis, though our choices will constrain the signs of the remaining elements in V. \\

Let us make a choice for the signs of our known coefficients in the matrix and choose them all to be positive for simplicity. We are now left with a much smaller set of conditions:
\begin{equation}\begin{gathered} \sum_{i=2}^4v_{ji}=-\frac{1}{2}\,\,,\\ \sum_{i=2}^4v_{ji}^2=\frac{3}{4}\,\,\, \text{and}\\  \sum_{i=2}^4v_{ji}v_{ki}=\frac{1}{4}\,\,,\\j,k\in\{2,3,4\}\,\,\,\text{and}\,\,\, k\neq j\,\,\,.\end{gathered}\end{equation}
After a few choice rearrangements, these coefficients can be calculated numerically in Mathematica. This yields a unitary matrix,
\begin{equation}V=\frac{1}{2}\left(\begin{array}{cccc}1&1&1&1\\1&-1&-1&1\\1&-1&1&-1\\1&1&-1&-1\end{array}\right)\,,\end{equation}
up to exchanges of the bottom three rows, which arises  due to the fact the triplet arising in this representation may be ordered arbitrarily. There is also a degree of choice involved regarding the sign of the rows. However, this is again largely unimportant as the result would be equivalent. \\

If we apply this transformation to our original basis \(t_i\), we find that we have a singlet and a triplet in the new basis,
\begin{equation}\begin{gathered}t_{singlet}=t_1+t_2+t_3+t_4\\ t_{triplet}=\left(t_1-t_2-t_3+t_4,\,t_1-t_2+t_3-t_4\,,t_1+t_2-t_3-t_4\right)\,,\end{gathered}\end{equation}
and that our generators become block-diagonal:
\begin{equation}\begin{gathered}S'=\left(\begin{array}{cccc}1&0&0&0\\0&1&0&0\\0&0&-1&0\\0&0&0&-1\end{array}\right)\\
T'=\left(\begin{array}{cccc}1&0&0&0\\0&0&1&0\\0&0&0&1\\0&1&0&0\end{array}\right)\,.\end{gathered}\end{equation}
\subsubsection{List of Conditions}
\begin{equation}\begin{gathered}0)\,\,\,v_{ij}\in\mathbb{R}\forall i,j\\
\text{1-4})\,\,\,\sum_{j=1}^{4}v_{ij}^2=1\,\,\,i\in\left\{1,2,3,4\right\}\\
5)\,\,\,v_{11}v_{21}+v_{12}v_{22}+v_{13}v_{23}+v_{14}v_{24}=0\\
6)\,\,\,v_{11}v_{31}+v_{12}v_{32}+v_{13}v_{33}+v_{14}v_{34}=\\
7)\,\,\,v_{11}v_{41}+v_{12}v_{42}+v_{13}v_{43}+v_{14}v_{44}=0\\
8)\,\,\,v_{21}v_{31}+v_{22}v_{32}+v_{23}v_{33}+v_{24}v_{34}=0\\
9)\,\,\,v_{21}v_{41}+v_{22}v_{42}+v_{23}v_{43}+v_{24}v_{44}=0\\
10)\,\,\,v_{31}v_{41}+v_{32}v_{42}+v_{33}v_{43}+v_{34}v_{44}=0\\
11)\,\,\,2v_{12}v_{13}+2v_{11}v_{14}=1\\
12)\,\,\,v_{11}v_{24}+v_{12}v_{23}+v_{13}v_{22}+v_{14}v_{21}=0\\
13)\,\,\,v_{11}v_{34}+v_{12}v_{33}+v_{13}v_{32}+v_{14}v_{31}=0\\
14)\,\,\,v_{11}v_{44}+v_{12}v_{43}+v_{13}v_{42}+v_{14}v_{41}=0\\
15)\,\,\,v_{11}^2+v_{12}v_{13}+v_{12}v_{14}+v_{13}v_{14}=1\\
16)\,\,\,v_{11}v_{21}+v_{12}v_{24}+v_{13}v_{22}+v_{14}v_{23}=0\\
17)\,\,\,v_{11}v_{31}+v_{12}v_{34}+v_{13}v_{32}+v_{14}v_{33}=0\\
18)\,\,\,v_{11}v_{41}+v_{12}v_{44}+v_{13}v_{42}+v_{14}v_{43}=0\\
19)\,\,\,v_{11}v_{21}+v_{12}v_{23}+v_{13}v_{24}+v_{14}v_{22}=0\\
20)\,\,\,v_{11}v_{31}+v_{12}v_{33}+v_{13}v_{34}+v_{14}v_{32}=0\\
21)\,\,\,v_{11}v_{41}+v_{12}v_{43}+v_{13}v_{44}+v_{14}v_{42}=0\\
22)\,\,\,v_{21}^2+v_{22}v_{23}+v_{22}v_{24}+v_{23}v_{24}=0\\
23)\,\,\,v_{31}^2+v_{32}v_{33}+v_{32}v_{34}+v_{33}v_{34}=0\\
24)\,\,\,v_{41}^2+v_{42}v_{43}+v_{42}v_{44}+v_{43}v_{44}=0\\
 \end{gathered}\end{equation}

\section{Yukawa coupling algebra}\label{ykwalgebra}
Table \ref{table5} specifies all the allowed operators for the $N=0$ $SU(5)\times A_4\times U(1)$ model discussed in the main text.  Here we include the full algebra for calculation of the Yukawa matrices given in the text. All couplings must have zero \(t_5\) charge, respect R-symmetry and be \(A_4\) singlets. In the basis derived in Appendix \ref{a4basis}, we have the triplet product:
\[\begin{gathered}
3_a\times3_b =1+1'+1''+3_1+3_2\\
1=a_1b_2+a_2b_2+a_3b_3\\
1'=a_1b_2+\omega a_2b_2+\omega^2a_3b_3\\
1''=a_1b_2+\omega^2a_2b_2+\omega a_3b_3\\
3_1=(a_2b_3,\,a_3b_1,\,a_1b_2)^{\text{T}}\\
3_2=(a_3b_2,\,a_1b_3,\,a_2b_1)^{\text{T}}\end{gathered}\]
where \(3_a=(a_1,\,a_2,\,a_3)^{\text{T}}\) and \(3_b=(b_1,\,b_2,\,b_3)^{\text{T}}\).
\subsection{Top-type quarks}
The top-type quarks have four non-vanishing couplings, while the \(T\cdot T_3\cdot H_u\cdot\theta_a\cdot\theta_a\) and \(T\cdot T\cdot H_u\cdot\theta_a\cdot\theta_a\cdot\theta_b\) couplings vanishings due to the chosen vacuum expectations: \(\langle H_u\rangle=(v,0,0)^{\text{T}}\) and \(\langle\theta_a\rangle=(a,0,0)^{\text{T}}\).\\

The contribution to the heaviest generation self-interaction is due to the \(T_3\cdot T_3\cdot H_u\cdot\theta_a\) operator:
\begin{align*}(1\times1)\times(3\times3)&\rightarrow 1\times 1\\&\rightarrow1\\
(T_3\times T_3)\times H_u\times\theta_a&\rightarrow (T_3\times T_3) va 
\end{align*} We note that this is the lowest order operator in the top-type quarks, so should dominate the hierarchy.\\

The interaction between the third generation and the lighter two generations is determined by the \(T\cdot T_3\cdot H_u\cdot\theta_a\cdot\theta_b\) operator:
\begin{align*}
(1\times1)\times(3\times3)\times1&\rightarrow 1\times1\times1\\
&\rightarrow 1\\
T\times T_3\times H_u\times\theta_a\times\theta_b&\rightarrow vab
\end{align*}\\
The remaining, first-second generation operators give contributions, in brief:
\begin{align*}
T\times T\times H_u\times\theta_a\times(\theta_b)^2&\rightarrow vab^2\\
T\times T\times H_u\times(\theta_a)^3&\rightarrow va^3
\end{align*}
These will be subject to Rank Theorem arguments, so that only one of the generations directly gets a mass from the Yukawa interaction. However the remaining generation will gain a mass due to instantons and non-commutative fluxes, as in \cite{Cecotti:2009zf}\cite{Aparicio:2011jx}.

\subsection{Charged Leptons}
The charged Leptons and Bottom-type quarks come from the same operators in the GUT group, though in this exposition we shall work in terms of the Charged Leptons. The complication for Charged leptons is that the Left-handed doublet is an \(A_4\) triplet, while the right-handed singlets of the weak interaction are singlets of the monodromy group. There are a total of six contributions to the Yukawa matrix, with the third generation right-handed types being generated by two operators. \\

The operators giving mass to the interactions of the right-handed third generation are dominated by the tree level operator \(F\cdot H_d\cdot T_3\), which gives a contribution as:
\begin{align*}
3\times3\times 1&\rightarrow 1\times 1\rightarrow 1\\
F\times H_d\times T_3&\rightarrow y_1 \left(\begin{array}{ccc}0&0&v_1\\0&0&v_2\\0&0&v_3\end{array}\right)
\end{align*}
Clearly this should dominated the next order operator, however when we choose a vacuum expectation for the \(H_d\) field, we will have contributions from \(F\cdot H_d\cdot T_3\cdot\theta_d\):
\begin{align*}
3\times3\times3\times1&\rightarrow 3\times3\times1\rightarrow1\\
F\times H_d\times\theta_d\times T_3&\rightarrow \left(\begin{array}{ccc}0&0&y_2v_2d_3+y_3v_3d_2\\0&0&y_2v_3d_1+y_3v_1d_3\\0&0&y_2v_1d_2+y_3v_2d_1\end{array}\right)
\end{align*}
The generation of Yukawas for the lighter two generations comes, at leading order, from the operators \(F\cdot H_d\cdot T\cdot\theta_b\) and \(F\cdot H_d\cdot T\cdot\theta_a\):
\begin{align*}
F\times H_d\times T \times \theta_b&\rightarrow y_4 b\left(\begin{array}
{ccc}v_1&v_1&0\\v_2&v_2&0\\v_3&v_3&0
\end{array}\right)\\
F\times H_d\times T \times \theta_a&\rightarrow y_5 a\left(\begin{array}
{ccc}0&0&0\\v_3&v_3&0\\v_2&v_2&0
\end{array}\right)\,,
\end{align*}
where the vacuum expectations for \(\theta_a\) and \(\theta_b\) are as before. The next order of operator take the same form, but with corrections due to the flavon triplet, \(\theta_d\). 
\begin{align*}
F\times H_d\times T \times \theta_b\times \theta_d&\rightarrow \left(\begin{array}
{ccc}y_6v_2d_3+y_7v_3d_2&y_6v_2d_3+y_7v_3d_2&0\\
y_6v_3d_1+y_7v_1d_3&y_6v_3d_1+y_7v_1d_3&0\\
y_6v_1d_2+y_7v_2d_1&y_6v_1d_2+y_7v_2d_1&0
\end{array}\right)
\end{align*}

\begin{align*}
F\times H_d\times T \times \theta_a\times \theta_d&\rightarrow a\left(\begin{array}
{ccc}y_8v_1d_1+y_{10}v_2d_2+y_{11}v_3d_3&y_8v_1d_1+y_{10}v_2d_2+y_{11}v_3d_3&0\\
y_{12}v_1d_2&y_{12}v_1d_2&0\\
y_9v_1d_3&y_9v_1d_3&0
\end{array}\right)
\end{align*}

\subsection{Neutrinos}
The neutrino sector admits masses of both Dirac and Majorana types. In the \(A_4\) model, the right-handed neutrino is assigned to a matter curve constituting a singlet of the GUT group. However it is a triplet of the \(A_4\) family symmetry, which along with the \(SU(2)\) doublet will generate complicated structures under the group algebra.   

\subsubsection{Dirac Mass Terms}
The Dirac mass terms coupling left and right-handed neutrinos comes from a maximum of four operators. The leading order operators are \(\theta_c\cdot F\cdot H_u\cdot \theta_b\)  and \(\theta_c\cdot F\cdot H_u\cdot \theta_a\), where as we have already seen the GUT singlet flavons \(\theta_a\) and \(\theta_b\) are used to cancel \(t_5\) charges. The right-handed neutrino is presumed to live on the GUT singlet \(\theta_d\) . \\

The first of the operators, \(\theta_c\cdot F\cdot H_u\cdot \theta_b\), contributes via two channels: 
\begin{align*}
3\times 3\times 3\times 1&\rightarrow 3 \times 3_a\times 1\rightarrow 1\times 1\\
&\rightarrow \left(\begin{array}{c}c_1\\c_2\\c_3\end{array}\right)\times \left(\begin{array}{c}F_2v_3\\F_3v_1\\F_1v_2\end{array}\right)\times b\rightarrow y_8b\left(\begin{array}{ccc}0&0&v_2\\v_3&0&0\\0&v_1&0\end{array}\right)\\
3\times 3\times 3\times 1&\rightarrow 3 \times 3_b\times 1\rightarrow 1\times 1\\
&\rightarrow \left(\begin{array}{c}c_1\\c_2\\c_3\end{array}\right)\times \left(\begin{array}{c}F_3v_2\\F_1v_3\\F_2v_1\end{array}\right)\times b\rightarrow y_9b\left(\begin{array}{ccc}0&v_3&0\\0&0&v_1\\v_2&0&0\end{array}\right)\\
\text{With the VEV alignments}&\text{ \(\langle\theta_a\rangle=(a,0,0)^{\text{T}}\) and \(\langle H_u\rangle=(v,0,0)^{\text{T}}\), we have a total matrix for the operator:}\\
&\rightarrow \left(\begin{array}{ccc}0&0&0\\0&0&y_9bv\\0&y_8bv&0\end{array}\right)
\end{align*}
The second leading order operator, \(\theta_c\cdot F\cdot H_u\cdot\theta_a\), is more cimplicated  due to the presence of four \(A_4\) triplet fields. The simpelst contribution to the operator is:
\begin{align*}
(3\times3)\times(3\times3)&\rightarrow 1\times 1\\
&\rightarrow\left(\begin{array}{ccc}
y_1 (v_1 a_1+v_2 a_2+v_3 a_3)&0&0\\
0&y_1 (v_1 a_1+v_2 a_2+v_3 a_3)&0\\
0&0&y_1 (v_1 a_1+v_2 a_2+v_3 a_3)\end{array}\right)\,,
\end{align*}
which only contributes to the diagonal. This is accompanied by two similar operators in the way of:
\begin{align*}
(3\times3)\times(3\times3)&\rightarrow 1'\times 1''\\
&\rightarrow (c_1F_1+\omega c_2F_2+\omega^2 c_3F_3)\times(v_1a_1+\omega^2 v_2a_2+\omega v_3a_3)\\
&\rightarrow y_2\left(\begin{array}{ccc}
v_1a_1&0&0\\
0&v_2a_2&0\\
0&0&v_3a_3
\end{array}\right)\\
(3\times3)\times(3\times3)&\rightarrow 1''\times 1'\\
&\rightarrow (c_1F_1+\omega^2 c_2F_2+\omega c_3F_3)\times(v_1a_1+\omega v_2a_2+\omega^2 v_3a_3)\\
&\rightarrow y_3\left(\begin{array}{ccc}
v_1a_1&0&0\\
0&v_2a_2&0\\
0&0&v_3a_3
\end{array}\right)\,.
\end{align*}
The remaining contribtuions are the complicated four-triplet products. However, upon retaining to our previous vacuum expectation values, these will all vanish, leaving an overall matrix of:
\begin{align*}
&\rightarrow\left(\begin{array}{ccc}
y_0va&0&0\\
0&y_1va&0\\
0&0&y_1va
\end{array}\right)
\end{align*}
Where \(y_0=y_1+y_2+y_3\) as before. These contributions will produce a large mixing between the second and third generations, however they do not allow for mixing with the first generation.\\

Corrections from the next order operators will give a weaker mixing with the first generation. These correcting terms are \(\theta_c\cdot F\cdot H_u\cdot\theta_d\cdot\theta_b\) and \(\theta_c\cdot F\cdot H_u\cdot\theta_d\cdot\theta_a\), though we choose to only consider the first of these two operators, since the flavon \(\theta_a\) will generate a very complicated structure, hindering computations with little obvious benefit in terms of model building. The \(\theta_c\cdot F\cdot H_u\cdot\theta_d\cdot\theta_b\) operator has of diagonal contributions as:
\begin{align*}
(3\times3)\times(3\times3)\times1&\rightarrow 3_a\times 3_x\times 1\rightarrow1\\
\theta_c\times F\times H_u\times\theta_d\times\theta_b&\rightarrow\left(\begin{array}
{c}c_2F_3\\c_3F_1\\c_1F_2\end{array}\right)\times \left(\begin{array}{c}0\\0\\vd_2\end{array}\right)\times b\\
&\rightarrow \left(\begin{array}{ccc}0&0&0\\z_1vd_2b&0&0\\0&0&0\end{array}\right)\,.\end{align*}
This is mirrored by similar combinations from the other 3 triplet-triplet combinations allowed by the algebra. Overall, this gives:

\begin{align*}
&\rightarrow \left(\begin{array}{ccc}0&z_3vd_2b&z_2vd_3b\\z_1vd_2b&0&0\\z_4vd_3b&0&0\end{array}\right)\,.\end{align*}
Due to the choice of Higgs vacuum expectation, the diagonal contributions will only correct the first generation mass, giving a contribution to it \(\sim vd_1b\).

\subsubsection{Majorana operators}
The right-handed neutrinos are also given a mass by Majorana terms. These are as it transpires relatively simple. The leading order term \(\theta_c\cdot\theta_c\), gives a diagonal contribtuion:
\begin{align*}
3\times3&\rightarrow1\\
\theta_c\cdot\theta_c&\rightarrow M \text{I}_{3\times3}
\end{align*}
There may also be corrections to the off diagonal, due to operators such as \(\theta_c\cdot\theta_c\cdot\theta_d\). These yield:\begin{align*}
3\times3\times3&\rightarrow3\times3\rightarrow1\\
\theta_c\times\theta_c\times\theta_d&\rightarrow \left(\begin{array}
{ccc}0&d_3&d_2\\ d_3&0&d_1\\ d_2&d_1&0
\end{array}\right)\,,
\end{align*}
Higher orders of the flavon \(\theta_d\) are also permitted, but should be suppressed by the coupling.

\section{Flux  mechanism}

For completeness, we discribe here in a simple manner the flux mechanism introduced
to break symmetries and generate chirality. 

$\bullet$ 
We start with  the $U(1)_Y$-flux  inside of $SU(5)_{GUT}$. 
 
 The ${\bf 5}$'s and ${\bf 10}$'s  reside on matter curves $\Sigma_{5_i}, \Sigma_{10_j}$ while 
  are characterised by their defining equations. 
From the latter, we can deduce the corresponding homologies $\chi_i$ following the standard procedure. 
If we turn on a  $U(1)_Y$-flux  ${\cal F}_Y$,
we can  determine the  flux restrictions  on them which are expressed in terms of integers through the ``dot product''
\[  N_{Y_i}= {\cal F}_Y\cdot \chi_i\]
The flux is responsible for the $SU(5)$ breaking down to the Standard Model and this can happen in such a way that the
$U(1)_Y$ gauge boson remains massless~\cite{Beasley:2008dc,Donagi:2008ca}. 
 On the other hand,  flux affects  the multiplicities of the SM-representations 
carrying non-zero $U(1)_Y$-charge.

Thus, on a certain  $\Sigma_{5_i}$ matter curve for example, we have 
 \ba {\bf 5}\in
SU(5)\Rightarrow\left\{\begin{array}{lll}
n_{(3,1)_{-\frac 13}}-n_{(\bar 3,1)_{\frac 13}}&=&{ M_5}\\
n_{(1,2)_{\frac 12}}-n_{(1, 2)_{-\frac 12}}&=&{ M_{5}}+ N_{Y_i}
\end{array}
\right. \label{5dec}
\ea
where $N_{Y_i}= {\cal F}_Y\cdot \chi_i$ as above.
We can arrange for example $M_5+N_{Y_i}=0$ to eliminate the doublets or $M_5=0$ 
to eliminate the triplet.

$\bullet$
Let's turn now to the $SU(5)\times S_3$.  The $S_3$ factor is associated to the three roots $t_{1,2,3}$ 
which can split to a singlet and a doublet
\[{\bf 1}_{S_3}=t_s=t_1+t_2+t_3,\; {\bf 2}_{S_3}=\{t_1-t_2,\; t_1+t_2-2 t_3\}^T\]
 It is convenient to introduce the two new linear combinations  
\[ t_a=t_1-t_3,\; t_b=t_2-t_3\]
and  rewrite the doublet as  follows
\be
 {\bf 2}_{S_3}=\left(\begin{array}{l}
t_a-t_b\\
t_a+t_b
\end{array}
\right)\;\to\; \left(\begin{array}{l}
-t_b\\
+t_b
\end{array}
\right)_{t_a}
\ee
Under the whole symmetry the $SU(5)_{GUT}$ ${\bf 10}_{t_i}, i=1,2,3$ representations transform
\[  ( {\bf 10}, {\bf 1}_{S_3})+ ( {\bf 10}, {\bf 2}_{S_3})\]

Our  intention is to turn on fluxes along certain directions. 
We can think of the following two different choices:

1) We can  turn on   a flux $N_a$ along $t_a$\footnote{In the
old basis we would require $N_{t_1}=\frac 23N_a$ and $N_{t_2}= N_{t_3}=-\frac 13N_a$.}. 
 The singlet $({\bf 10}, {\bf 1}_{S_3})$ does not transform 
under $t_a$,   hence this flux will  split the  
multiplicities as follows
 \ba 
{\bf 10}_{t_i} \Rightarrow\left\{\begin{array}{lll}
( {\bf 10}, {\bf 1}_{S_3})&=&{ M}\\
( {\bf 10}, {\bf 2}_{S_3})&=&{ M}+ N_{a}
\end{array}
\right.
\ea
This choice will also break the $S_3$ symmetry to $Z_3$.

2) Turning on  a flux along the singlet direction $t_s$ will preserve $S_3$ symmetry. 
The multiplicities now read
 \ba 
{\bf 10}_{t_i} \Rightarrow\left\{\begin{array}{lll}
( {\bf 10}, {\bf 1}_{S_3})&=&{ M}+ N_{s}\\
( {\bf 10}, {\bf 2}_{S_3})&=&{ M}
\end{array}
\right.
\ea
To get rid of the doublets we choose $M=0$  while because flux restricts  non-trivially
on the matter curve, the number of singlets can differ by just choosing   $ N_{s}\ne 0$.

\newpage

\section{The $b_1=0$ constraint}

To solve  the $b_1=0$ constraint we have repeatidly introduced a new section $a_0$ and 
assumed factorisation of the involved $a_i$ coefficients. To check the validity of this
assumption, we take as an example the $S_3\times Z_2$ case, where $b_1=a_2a_6+a_3a_5=0$. 
We note first that the coefficients $b_k$ are holomorphic   functions  of $z$,   
and as such they can be expressed as power series of the form $b_k = b_{k,0}+b_{k,1}z+\cdots$ 
where $b_{k,m}$ do not depend on $z$. Hence, the coefficients $a_k$  have  a $z$-independent part
\[ a_k = \sum_{m=0}a_{k,m}z^m\]
while the product of two of them can be cast to the form
\[a_l\, a_k = \sum_{p=0}\beta_p\,z^p,\;\; {\rm with }\;\; \beta_p=\sum_{n=0}^pa_{ln}a_{k,p-n}\]

Clearly the condition $b_1=a_2a_6+a_3a_5=0$ has to be satisfied term-by-term. To this end, at
the next to zeroth order we  define
\be
\lambda = \frac{a_{3,1}a_{5,0}+a_{2,1}a_{6,0}}{a_{5,1}a_{6,0}-a_{5,0}a_{6,1}}\label{lambdadef}
\ee
  The requirement  $a_{5,1}a_{6,0}\ne a_{5,0}a_{6,1}$
ensures  finiteness of  $\lambda$,  while at the same time 
excludes a relation of the form   $ a_5\propto \kappa a_6$ 
where $\kappa$ would be a new  section.

We can write the expansions for $a_2,a_3$ as follows
\be
\
\begin{split}
a_2&=\lambda a_{5,0}+a_{2,1}z+{\cal O}(z^2)\\
a_3&=-\lambda a_{6,0}+a_{3,1}z+{\cal O}(z^2)
\end{split}
\ee
The $b_1=0$ condition is now
\[ b_1= 0+0\, z +{\cal O}(z^2)\]
i.e., satified up to second order in $z$.  Hence, locally we can set $z=0$ and simply write
\[ a_2=\lambda \,a_5,\;  a_3=-\lambda\, a_6\]


\newpage


\begin{thebibliography}{99}

\bibitem{Vafa:1996xn}
  C.~Vafa,
  Nucl.\ Phys.\  B {\bf 469} (1996) 403
  [arXiv:hep-th/9602022].


        \bibitem{Donagi:2008ca}
          R.~Donagi and M.~Wijnholt,
          ``Model Building with F-Theory,''
          Adv.\ Theor.\ Math.\ Phys.\  {\bf 15} (2011) 1237
          [arXiv:0802.2969 [hep-th]].



\bibitem{Beasley:2008dc}
  C.~Beasley, J.~J.~Heckman and C.~Vafa,
  ``GUTs and Exceptional Branes in F-theory - I,''
  JHEP {\bf 0901} (2009) 058
  [arXiv:0802.3391 [hep-th]].

 



  \bibitem{Beasley:2008kw}
  C.~Beasley, J.~J.~Heckman and C.~Vafa,
  JHEP {\bf 0901} (2009) 059
  [arXiv:0806.0102 [hep-th]].

\bibitem{Blumenhagen:2009yv}
  R.~Blumenhagen, T.~W.~Grimm, B.~Jurke and T.~Weigand,
  Nucl.\ Phys.\  B {\bf 829} (2010) 325
  [arXiv:0908.1784].

\bibitem{Heckman:2008qa}
  J.~J.~Heckman and C.~Vafa,
  Nucl.\ Phys.\  B {\bf 837} (2010) 137
  [arXiv:0811.2417 [hep-th]].

\bibitem{Heckman:2008es}
  J.~J.~Heckman, J.~Marsano, N.~Saulina, S.~Schafer-Nameki and C.~Vafa,
  arXiv:0808.1286 [hep-th].

                 \bibitem{Blumenhagen:2008zz}
                   R.~Blumenhagen, V.~Braun, T.~W.~Grimm and T.~Weigand,
                   Nucl.\ Phys.\ B {\bf 815} (2009) 1
                   [arXiv:0811.2936 [hep-th]].



\bibitem{Denef:2008wq}
  F.~Denef,
  arXiv:0803.1194.

\bibitem{Weigand:2010wm}
  T.~Weigand,
  Class.\ Quant.\ Grav.\  {\bf 27} (2010) 214004
  [arXiv:1009.3497 [hep-th]].

\bibitem{Heckman:2010bq}
  J.~J.~Heckman,
  Ann.\ Rev.\ Nucl.\ Part.\ Sci.\  {\bf 60} (2010) 237
  [arXiv:1001.0577 [hep-th]].

\bibitem{Grimm:2010ks}
  T.~W.~Grimm,
  Nucl.\ Phys.\  B {\bf 845} (2011) 48
  [arXiv:1008.4133 [hep-th]].
  
  \bibitem{Leontaris:2012mh}
    G.~K.~Leontaris,
    PoS CORFU {\bf 2011} (2011) 095
    [arXiv:1203.6277 [hep-th]].
    
    \bibitem{Maharana:2012tu}
      A.~Maharana and E.~Palti,
      Int.\ J.\ Mod.\ Phys.\ A {\bf 28} (2013) 1330005
      [arXiv:1212.0555 [hep-th]].



\bibitem{Heckman:2009mn}
  J.~J.~Heckman, A.~Tavanfar and C.~Vafa,
  JHEP {\bf 1008} (2010) 040
  [arXiv:0906.0581 [hep-th]].

    
                          

    



\bibitem{Donagi:2008kj}
  R.~Donagi and M.~Wijnholt,
  Adv.\ Theor.\ Math.\ Phys.\  {\bf 15} (2011) 1523
  [arXiv:0808.2223 [hep-th]].

  
  
    \bibitem{Marsano:2009gv}
      J.~Marsano, N.~Saulina and S.~Schafer-Nameki,
      ``Monodromies, Fluxes, and Compact Three-Generation F-theory GUTs,''
      JHEP {\bf 0908} (2009) 046
      [arXiv:0906.4672 [hep-th]].
      
      
        \bibitem{Hayashi:2009bt}
          H.~Hayashi, T.~Kawano, Y.~Tsuchiya and T.~Watari,
          ``Flavor Structure in F-theory Compactifications,''
          JHEP {\bf 1008} (2010) 036
          [arXiv:0910.2762 [hep-th]].
        
              \bibitem{Dudas:2010zb}
                E.~Dudas and E.~Palti,
                ``On hypercharge flux and exotics in F-theory GUTs,''
                JHEP {\bf 1009} (2010) 013
                [arXiv:1007.1297 [hep-ph]].
  
  
  
      \bibitem{King:2010mq}
                  S.~F.~King, G.~K.~Leontaris and G.~G.~Ross,
                  ``Family symmetries in F-theory GUTs,''
                  Nucl.\ Phys.\ B {\bf 838} (2010) 119
                  [arXiv:1005.1025 [hep-ph]].
                
   
   
    \bibitem{Callaghan:2011jj}
      J.~C.~Callaghan, S.~F.~King, G.~K.~Leontaris and G.~G.~Ross,
      ``Towards a Realistic F-theory GUT,''
      JHEP {\bf 1204} (2012) 094
      [arXiv:1109.1399 [hep-ph]].
      
      
      
      
            \bibitem{Antoniadis:2012yk}
        I.~Antoniadis and G.~K.~Leontaris,
        ``Building SO(10) models from F-theory,''
        JHEP {\bf 1208} (2012) 001
        [arXiv:1205.6930 [hep-th]].
          
        
      \bibitem{Callaghan:2012rv}
        J.~C.~Callaghan and S.~F.~King,
        ``E6 Models from F-theory,''
        JHEP {\bf 1304} (2013) 034
        [arXiv:1210.6913 [hep-ph]].
        
        
        \bibitem{Tatar:2012tm}
          R.~Tatar and W.~Walters,
          ``GUT theories from Calabi-Yau 4-folds with SO(10) Singularities,''
          JHEP {\bf 1212} (2012) 092
          [arXiv:1206.5090 [hep-th]].

\bibitem{Callaghan:2013kaa}
  J.~C.~Callaghan, S.~F.~King and G.~K.~Leontaris,
  JHEP {\bf 1312} (2013) 037
  [arXiv:1307.4593 [hep-ph]].
  
  
  
  \bibitem{Ibanez:1991hv}
    L.~E.~Ibanez and G.~G.~Ross,
    ``Discrete gauge symmetry anomalies,''
    Phys.\ Lett.\ B {\bf 260} (1991) 291.
  
  \bibitem{Anastasopoulos:2012zu}
    P.~Anastasopoulos, M.~Cvetic, R.~Richter and P.~K.~S.~Vaudrevange,
    ``String Constraints on Discrete Symmetries in MSSM Type II Quivers,''
    JHEP {\bf 1303} (2013) 011
    [arXiv:1211.1017 [hep-th]].
    
  \bibitem{Lee:2010gv}
    H.~M.~Lee, S.~Raby, M.~Ratz, G.~G.~Ross, R.~Schieren, K.~Schmidt-Hoberg and P.~K.~S.~Vaudrevange,
    ``A unique $Z_4^R$ symmetry for the MSSM,''
    Phys.\ Lett.\ B {\bf 694} (2011) 491
    [arXiv:1009.0905 [hep-ph]].
    
  \bibitem{Ibanez:2012wg}
    L.~E.~Ibanez, A.~N.~Schellekens and A.~M.~Uranga,
    ``Discrete Gauge Symmetries in Discrete MSSM-like Orientifolds,''
    Nucl.\ Phys.\ B {\bf 865} (2012) 509
    [arXiv:1205.5364 [hep-th]].
    
    \bibitem{Honecker:2013hda}
      G.~Honecker and W.~Staessens,
      ``To Tilt or Not To Tilt: Discrete Gauge Symmetries in Global Intersecting D-Brane Models,''
      JHEP {\bf 1310} (2013) 146
      [arXiv:1303.4415 [hep-th]].
  
  \bibitem{BerasaluceGonzalez:2012vb}
    M.~Berasaluce-Gonzalez, P.~G.~Camara, F.~Marchesano, D.~Regalado and A.~M.~Uranga,
    ``Non-Abelian discrete gauge symmetries in 4d string models,''
    JHEP {\bf 1209} (2012) 059
    [arXiv:1206.2383 [hep-th]].
  
  

  
  \bibitem{Altarelli:2010gt}
    G.~Altarelli and F.~Feruglio,
    ``Discrete Flavor Symmetries and Models of Neutrino Mixing,''
    Rev.\ Mod.\ Phys.\  {\bf 82} (2010) 2701
    [arXiv:1002.0211 [hep-ph]].
    

\bibitem{King:2013eh}
  S.~F.~King and C.~Luhn,
  ``Neutrino Mass and Mixing with Discrete Symmetry,''
  Rept.\ Prog.\ Phys.\  {\bf 76} (2013) 056201
  [arXiv:1301.1340 [hep-ph]].

\bibitem{King:2014nza}
  S.~F.~King, A.~Merle, S.~Morisi, Y.~Shimizu and M.~Tanimoto,
  New J.\ Phys.\  {\bf 16} (2014) 045018
  [arXiv:1402.4271 [hep-ph]].
  
  
  
  
  






  
  


     
         
                                                          
                           
                                                                
                                                      
                                               

    
           
         
                                 
                          
                          
                          \bibitem{Antoniadis:2013joa}
                            I.~Antoniadis and G.~K.~Leontaris,
                            ``Neutrino mass textures from F-theory,''
                            Eur.\ Phys.\ J.\ C {\bf 73} (2013) 2670
                            [arXiv:1308.1581 [hep-th]].
                            
                          
                          
                        \bibitem{Donagi:2009ra}
                          R.~Donagi and M.~Wijnholt,
                          Commun.\ Math.\ Phys.\  {\bf 326} (2014) 287
                          [arXiv:0904.1218 [hep-th]].
                          
                         
                         
                          \bibitem{Cecotti:2009zf}
                            S.~Cecotti, M.~C.~N.~Cheng, J.~J.~Heckman and C.~Vafa,
                            arXiv:0910.0477 [hep-th].
                            
                            
                                                    \bibitem{Aparicio:2011jx}
                            L.~Aparicio, A.~Font, L.~E.~Ibanez and F.~Marchesano,
                            JHEP {\bf 1108} (2011) 152
                            [arXiv:1104.2609 [hep-th]].
                          \bibitem{marchesano}
                            F.~Marchesano and L.~Martucci,
                            Phys.\ Rev.\ Lett.\  {\bf 104} (2010) 231601
                            [arXiv:0910.5496 [hep-th]].
                            
                            \bibitem{Font:2013ida}
                              A.~Font, F.~Marchesano, D.~Regalado and G.~Zoccarato,
                              JHEP {\bf 1311} (2013) 125
                              [arXiv:1307.8089 [hep-th]].
                            
                       
                          
                          
                          
                          \bibitem{Mayrhofer:2013ara}
                             C.~Mayrhofer, E.~Palti and T.~Weigand,
                             JHEP {\bf 1309} (2013) 082
                             [arXiv:1303.3589 [hep-th]].
              
  \bibitem{King:2003jb}
  S.~F.~King,
  Rept.\ Prog.\ Phys.\  {\bf 67} (2004) 107
  [hep-ph/0310204].
  
  
  
                
    \bibitem{Nakayama88}
    N.     {Nakayama}, ``On Weierstrass models'', Algebraic Geometry and Commutative Algebra, Kinokuniya, Tokyo 1988
    
    
    
    
     \bibitem{Bouchard:2009bu}
       V.~Bouchard, J.~J.~Heckman, J.~Seo and C.~Vafa,
       JHEP {\bf 1001} (2010) 061
       [arXiv:0904.1419 [hep-ph]].


\bibitem{GonzalezGarcia:2012sz}
  M.~C.~Gonzalez-Garcia, M.~Maltoni, J.~Salvado and T.~Schwetz,
  JHEP {\bf 1212} (2012) 123
  [arXiv:1209.3023 [hep-ph]].

\bibitem{Lesgourgues:2012uu}
  J.~Lesgourgues and S.~Pastor,
  Adv.\ High Energy Phys.\  {\bf 2012} (2012) 608515
  [arXiv:1212.6154 [hep-ph]].

\bibitem{Latimer:2004hd}
  D.~C.~Latimer and D.~J.~Ernst,
  Phys.\ Rev.\ D {\bf 71} (2005) 017301
  [nucl-th/0405073].

\bibitem{Huber:2011wv}
  P.~Huber,
  Phys.\ Rev.\ C {\bf 84} (2011) 024617
   [Erratum-ibid.\ C {\bf 85} (2012) 029901]
  [arXiv:1106.0687 [hep-ph]].

\bibitem{Leontaris:2010zd}
  G.~K.~Leontaris and G.~G.~Ross,
  JHEP {\bf 1102} (2011) 108
  [arXiv:1009.6000 [hep-th]].
  
        \bibitem{Mayrhofer:2012zy}
          C.~Mayrhofer, E.~Palti and T.~Weigand,
          ``U(1) symmetries in F-theory GUTs with multiple sections,''
          JHEP {\bf 1303} (2013) 098
          [arXiv:1211.6742 [hep-th]].
   
          
          
          \bibitem{Borchmann:2013jwa}
            J.~Borchmann, C.~Mayrhofer, E.~Palti and T.~Weigand,
            Phys.\ Rev.\ D {\bf 88} (2013) 046005
            [arXiv:1303.5054 [hep-th]].
            
            \bibitem{Braun:2013nqa}
              V.~Braun, T.~W.~Grimm and J.~Keitel,
              JHEP {\bf 1312} (2013) 069
              [arXiv:1306.0577 [hep-th]].
              
              \bibitem{Cvetic:2013nia}
                M.~Cvetic, D.~Klevers and H.~Piragua,
                JHEP {\bf 1306} (2013) 067
                [arXiv:1303.6970 [hep-th]].
                
                \bibitem{Cvetic:2013qsa}
                  M.~Cvetic, D.~Klevers, H.~Piragua and P.~Song,
                  arXiv:1310.0463 [hep-th].
                
\bibitem{Marsano:2010sq}
  J.~Marsano,
  Phys.\ Rev.\ Lett.\  {\bf 106} (2011) 081601
  [arXiv:1011.2212 [hep-th]].
                     
                        
                        \bibitem{Krippendorf:2014xba}
                          S.~Krippendorf, D.~K.~M.~Pena, P.~-K.~Oehlmann and F.~Ruehle,
                          arXiv:1401.5084 [hep-th].
\bibitem{Ade:2013zuv}
  P.~A.~R.~Ade {\it et al.}  [Planck Collaboration],
  arXiv:1303.5076 [astro-ph.CO].

\bibitem{Font:2012wq}
  A.~Font, L.~E.~Ibanez, F.~Marchesano and D.~Regalado,
  JHEP {\bf 1303} (2013) 140
   [Erratum-ibid.\  {\bf 1307} (2013) 036]
  [arXiv:1211.6529 [hep-th]].


\end{thebibliography}
\end{document}